\documentclass[twocolumn]{aastex63}
\usepackage{xcolor,colortbl}
\usepackage[referable]{threeparttablex}

\usepackage{amsmath}
\newcommand*\subtxt[1]{_{\textnormal{#1}}}
\DeclareRobustCommand\_{\ifmmode\expandafter\subtxt\else\textunderscore\fi}

\graphicspath{{./}{figures/}}

\shorttitle{Characterization of the L 98-59 planets}
\shortauthors{Pidhorodetska et al.}

\usepackage{float}
\usepackage{multirow}
\begin{document}

\title{L~98-59: a Benchmark System of Small Planets for Future Atmospheric Characterization}

\author[0000-0001-9771-7953]{Daria Pidhorodetska}
\affiliation{Department of Earth and Planetary Sciences, University of California, Riverside, CA, USA}

\author[0000-0002-6721-3284]{Sarah E. Moran}
\affil{Department of Earth and Planetary Sciences,
 Johns Hopkins University, 3400 N. Charles Street,
Baltimore, MD 21218, USA}

\author[0000-0002-2949-2163]{Edward W. Schwieterman}
\affiliation{Department of Earth and Planetary Sciences, University of California, Riverside, CA, USA}
\affiliation{NASA Astrobiology Alternative Earths Team, Riverside, CA, USA}
\affiliation{Nexus for Exoplanet System Science (NExSS) Virtual Planetary Laboratory, Seattle, WA, USA}
\affiliation{Blue Marble Space Institute of Science, Seattle, WA, USA}

\author[0000-0001-7139-2724]{Thomas Barclay}
\affiliation{NASA Goddard Space Flight Center, 8800 Greenbelt Road, Greenbelt, MD 20771, USA}
\affiliation{University of Maryland, Baltimore County, 1000 Hilltop Cir, Baltimore, MD 21250, USA}

\author[0000-0002-5967-9631]{Thomas J. Fauchez}
\affiliation{NASA Goddard Space Flight Center,
8800 Greenbelt Road, Greenbelt, MD 20771, USA}
\affiliation{Goddard Earth Sciences Technology and Research (GESTAR), Universities Space Research Association (USRA), Columbia, MD}
\affiliation{GSFC Sellers Exoplanet Environments Collaboration}

\author[0000-0002-8507-1304]{Nikole K. Lewis}
\affiliation{Department of Astronomy and Carl Sagan Institute,
Cornell University, 122 Sciences Drive, Ithaca, NY 14853, USA}

\author[0000-0003-1309-2904]{Elisa V. Quintana}
\affiliation{NASA Goddard Space Flight Center, 8800 Greenbelt Road, Greenbelt, MD 20771, USA}

\author[0000-0002-2662-5776]{Geronimo L. Villanueva}
\affiliation{NASA Goddard Space Flight Center,
8800 Greenbelt Road,
Greenbelt, MD 20771, USA}

\author[0000-0003-0354-9325]{Shawn D. Domagal-Goldman}
\affiliation{Nexus for Exoplanet System Science (NExSS) Virtual Planetary Laboratory, Seattle, WA, USA}
\affiliation{NASA Goddard Space Flight Center,
8800 Greenbelt Road,
Greenbelt, MD 20771, USA}

\author[0000-0001-5347-7062]{Joshua E. Schlieder}
\affiliation{NASA Goddard Space Flight Center,
8800 Greenbelt Road,
Greenbelt, MD 20771, USA}

\author[0000-0002-0388-8004]{Emily A. Gilbert}
\affiliation{Department of Astronomy and Astrophysics, University of
Chicago, 5640 S. Ellis Ave, Chicago, IL 60637, USA}
\affiliation{University of Maryland, Baltimore County, 1000 Hilltop Circle, Baltimore, MD 21250, USA}
\affiliation{The Adler Planetarium, 1300 South Lakeshore Drive, Chicago, IL 60605, USA}
\affiliation{NASA Goddard Space Flight Center, 8800 Greenbelt Road, Greenbelt, MD 20771, USA}
\affiliation{GSFC Sellers Exoplanet Environments Collaboration}

\author[0000-0002-7084-0529]{Stephen R. Kane}
\affiliation{Department of Earth and Planetary Sciences, University of California, Riverside, CA, USA}

\author[0000-0001-9786-1031]{Veselin B. Kostov}
\affiliation{NASA Goddard Space Flight Center, 8800 Greenbelt Road, Greenbelt, MD 20771, USA}
\affiliation{SETI Institute, 189 Bernardo Ave, Suite 200, Mountain View, CA 94043, USA}
\affiliation{GSFC Sellers Exoplanet Environments Collaboration}

\begin{abstract}
L~98-59 is an M3V dwarf star that hosts three small (R $<$ 1.6 R$_\oplus$) planets. The host star is bright (K = 7.1) and nearby (10.6 pc), making the system a prime target for follow-up characterization with the Hubble Space Telescope (HST) and the upcoming James Webb Space Telescope (JWST). Herein, we use simulated transmission spectroscopy to evaluate the detectability of spectral features with HST and JWST assuming diverse atmospheric scenarios (e.g., atmospheres dominated by H$_2$, H$_2$O, CO$_2$, or O$_2$). We find that H$_2$O and CH$_4$ present in a low mean-molecular weight atmosphere could be detected with HST in 1 transit for the two outermost planets, while H$_2$O in a clear steam atmosphere could be detected in 6 transits or fewer with HST for all three planets. We predict that observations using JWST/NIRISS would be capable of detecting a clear steam atmosphere in 1 transit for each planet, and H$_2$O absorption in a hazy steam atmosphere in 2 transits or less. In a clear, desiccated atmosphere, O$_2$ absorption may be detectable for all three planets with NIRISS. If the L 98-59 planets possess a clear, Venus-like atmosphere, NIRSpec could detect CO$_2$ within 26 transits for each planet, but the presence of H$_2$SO$_4$ clouds would significantly suppress CO$_2$ absorption. The L 98-59 system is an excellent laboratory for comparative planetary studies of transiting multiplanet systems, and observations of the system via HST and JWST would present a unique opportunity to test the accuracy of the models presented in this study.
\end{abstract} 


\section{Introduction} \label{sec:intro}
Thousands of planets orbiting stars outside of our solar system have been confirmed, the vast majority of which were detected with the transit method. The Kepler and K2 missions \citep{Borucki2010,howell2014} taught us that planets and their host stars are remarkably diverse \citep[e.g.][]{Lissauer2011,Borucki2011,Batalha2014,Gaudi2020,Bryson2020}, and that planets are likely more abundant than stars in our galaxy \citep{Burke2015,Dressing2015}. While Kepler provided an unprecedented bounty of planets that enabled exoplanet population statistics, relatively few planets discovered by Kepler make ideal follow-up targets due to their large average distances. However, recent discoveries by ground-based telescopes of more readily characterizable Earth-sized worlds, such as the seven TRAPPIST-1 planets \citep{Gillon2016, Gillon2017}, are enabling us to learn about the properties of rocky planets beyond our solar system. The \text{Transiting Exoplanet Survey Satellite} \citep[\text{TESS};][]{Ricker2015}, launched in 2018, is currently performing a near-all-sky survey to search for planets orbiting bright stars in our solar neighborhood. Planets that orbit bright stars are favorable targets to probe their bulk compositions and atmospheres, particularly those orbiting M dwarfs because of the large planet-to-star radius ratio, and favorable transit probabilities for close-in planets. As TESS reveals new exoplanets, the prospect of discovering new worlds and probing their atmospheres provides an exciting opportunity to broaden our understanding of the nature of planets that may resemble our own. \\ 

Despite having a shorter observing baseline than Kepler, TESS has begun to uncover several multi-planet systems. Transiting multiplanet systems provide controlled environments for comparative planetary studies, as planets that form around the same star - and thus come from the same nebula and protoplanetary disk - share many properties that are typically less well constrained when comparing planets in different systems. Such shared properties include the host star's mass, composition, activity level, formation history, and evolution. Multiple planet systems in resonant chains also strongly increase the exchange of torque at each planet conjunction, offering a unique opportunity to characterize the physical properties through transit timing variations (TTV) \citep{Holman2005,agol2005}. However, the vast majority of known multiplanet systems (transiting or otherwise) orbit stars that are too distant and too faint for atmospheric characterization studies with the  Hubble Space Telescope (HST) or the upcoming James Webb Space Telescope (JWST). \\

Small (R $<$ 1.6 R$_\oplus$), potentially-rocky planets that are suitable for such studies are particularly scarce. To date, the only small (sub-1.6 R$_\oplus$)
planets with transmission spectroscopy measurements are those in the TRAPPIST-1 system \citep{deWit2016,deWit2018,Gillon2017,Luger2017,Delrez2018,Burdanov2019,Ducrot2018,Ducrot2020,Agol2021}, and the single-planet system GJ~1132~b \citep{bertathompson2015,southworth2017,DiamondLowe2018}. While no significant atmospheric detection has yet been made for any of these worlds, these observations have been highly informative. For example, HST observations of the TRAPPIST-1 planets have ruled out cloud/haze-free H$_2$-dominated atmospheres \citep{deWit2016,deWit2018,Wakeford2019}. \citet{Moran2018} explored whether hazy H$_2$-rich atmospheres could explain HST observations, and found that laboratory measurements suggested that haze formation would be inefficient for hot H$_2$ atmospheres. \citet{deWit2016,deWit2018}, \citet{Moran2018}, and \citet{Wakeford2019} also determined that these planets' atmospheres may be composed of a wide variety of compositions dominated by higher molecular weight species such as  N$_2$, O$_2$, H$_2$O, CO$_2$, or CH$_4$. As important as these results are, due to the limited sample size it may be premature to draw general conclusions. Fortunately, TESS is eagerly anticipated to discover a few benchmark systems, with nearby bright stars, large planet-to-star ratios, and close-in planets that will provide numerous transits that amplify their signal \citep{Barclay2018}. \\

Many of the planets that are the focus of follow-up observations, including atmospheric characterization, orbit M-dwarfs. Such planets are important for constraining these observations, as low-mass stars account for the majority of stars in the galaxy \citep{Henry2006,Winters2015}. M-dwarfs range from about a tenth to half of a solar mass and are significantly less luminous than the Sun (luminosities range from one twentieth of the Sun's luminosity from M0V stars to as low as several thousand times less luminous than the Sun for late M-dwarfs), but display variable but most often higher activity levels \citep{Kiraga2011,Hawley2014}. M-dwarf stars undergo a prolonged, high luminosity pre-main-sequence phase \citep{Ramirez2014,Tian2015,Luger2015}, where the total luminosity of the star can be as much as two orders of magnitude larger than the main-sequence luminosity. This has the potential to desiccate a planet, by evaporating all/any water from the surface to the atmosphere, where strong UV radiation from the star can photodissociate water-molecules into hydrogen and oxygen, creating the radicals H and OH (in some cases O). Hydrogen can escape to space, potentially leaving oxygen behind. This process would also catalyze the destruction and reprocessing of many gases in the atmosphere, reducing their overall lifetime (e.g., CO + OH $\rightarrow$ CO$_2$ + H). In turn, M-dwarf planets can retain their atmospheres from this phase, either through outgassing of secondary atmospheres, or surviving the pre-main-sequence phase due to large initial volatile endowments \citep{Bolmont2017,Bourrier2017}. \\

Multiplanet systems around M dwarf stars are common \citep[e.g.][]{Howard2012,Muirhead2015,Mulders2015,HardegreeUllman2019,Hsu2020}. With numerous small planets in multiplanet systems, we can delve into questions about the origin and evolution of such small planets. Such queries can be solved with transmission photometry and spectroscopy -- measuring the starlight filtered through the planet's atmosphere and observing the absorption at particular wavelengths to infer the presence of specific atoms and molecules. In making these measurements, we naturally must ask: what kind of atmospheres do these worlds have? Where did these atmospheres come from? Are they primordial and hydrogen-rich, as with comparatively well characterized hot Jupiters, or are they secondary outgassed atmospheres primarily composed of heavier molecules such as H$_2$O, CO$_2$, or even O$_2$? How do their atmospheric compositions vary depending on their current incident radiation? Do they contain clouds or hazes? Do rocky planets orbiting M dwarfs have atmospheres at all? A particularly intriguing planetary system orbits the nearby M-dwarf L 98-59.\\

Three planets orbiting the nearby star L~98-59 were detected by the TESS transit detection pipeline \citep{Jenkins2016} and their planetary nature confirmed through supporting ground-based observations and statistical analyses \citep{Kostov2019}. L~98-59 is an M3V star with M$_{*}$ = 0.31 $\pm$ 0.03 M$_\sun$, R$_{*}$ = 0.31 $\pm$ 0.01 R$_\sun$, and T$\subtxt{eff}$ = 3412 $\pm$ 49 K \citep{cloutier2019}. L~98-59 is a Main Sequence M dwarf and estimated to be $>$1 Gyr in age \citep{Kostov2019b}, inferred from the star's HR diagram position, lack of spectroscopic youth indicators, slow rotation, no evidence of variability due to spots, and low levels of white-light flare activity as seen in TESS data. This quiescence indicates that the star is beyond the active youth phase characteristic of young M dwarfs. The lack of spot variability and infrequent flare rate is advantageous for conducting transit spectroscopy observations of the planets in this system.\\ 

At 10.6 pc, L~98-59 is the second closest transiting multiplanet system to the Sun (after HD 219134). The planets have orbital periods of 2.25, 3.69, and 7.45 days, encompassing a near-resonant configuration. The masses of the outer two planets have been measured using ground-based radial velocity observations from HARPS \citep{cloutier2019}, at 2.42 $\pm$ 0.35 M$_\earth$ for L 98-59 c and 2.31 $\pm$ 0.46 M$_\earth$ for L 98-59 d. L 98-59 b is the smallest, innermost planet and the current RV measurements only allow for an upper limit of its mass at $<$1.01 M$_\earth$ (at 95\% confidence). These masses combined with the radii of the three planets (${\rm 0.80\ to\ 1.57\ R_{\oplus}}$) indicate that the two innermost planets have bulk densities that are almost certainly rocky \citep{Rogers2015,Dressing2015b,Chen_2016,Fulton2017,Owen2017}, while the outermost planet yields a bulk density that is consistent with substantial volatile content, suggesting that it is likely a mini-Neptune. \\

The L~98-59 planets receive significantly more energy than modern Earth receives from the Sun (a factor of 4--24 more than Earth's current irradiation), placing them in the Venus Zone \citep[VZ;][]{Kane2014}, an annulus where the received flux of the planets would likely push a terrestrial planet like Earth into a runaway greenhouse state, producing conditions similar to those found on Venus while not completely stripping the atmosphere of volatiles or refractory elements such as silicates. Studying planets that may be undergoing, or may have evolved into a post-runaway greenhouse state, is valuable as it can help teach us why some planets are Venus-like and others Earth-like \citep{Ehrenreich2012}. Although the post-runaway greenhouse state of Venus has rendered its atmosphere to be much different than that of Earth and ultimately uninhabitable, Venus and Earth share strong similarities in their size, density, and composition. Thus, studying Venus analogs allows us to place constraints on habitability, as atmospheric evolution of Venus/Earth-sized planets points towards runaway greenhouse conditions \citep{Kasting1988, Leconte2013b, Kane2018}. Although TESS is expected to discover $\sim$300 Venus-analog planets \citep{Ostberg2019}, most Venus-analog candidates discovered to date orbit relatively faint stars \citep{Barclay2013,Kane2013,Angelo2017,Gillon2017,Kane2018}. Fortunately, L 98-59 is bright (K = 7.1), and will therefore enable follow-up characterization of the planets' atmospheres, revealing which have Venus-like conditions. In that respect, L~98-59 could become a benchmark system. \\

\begin{figure} 
 \centering 
  \resizebox{9cm}{!}{\includegraphics{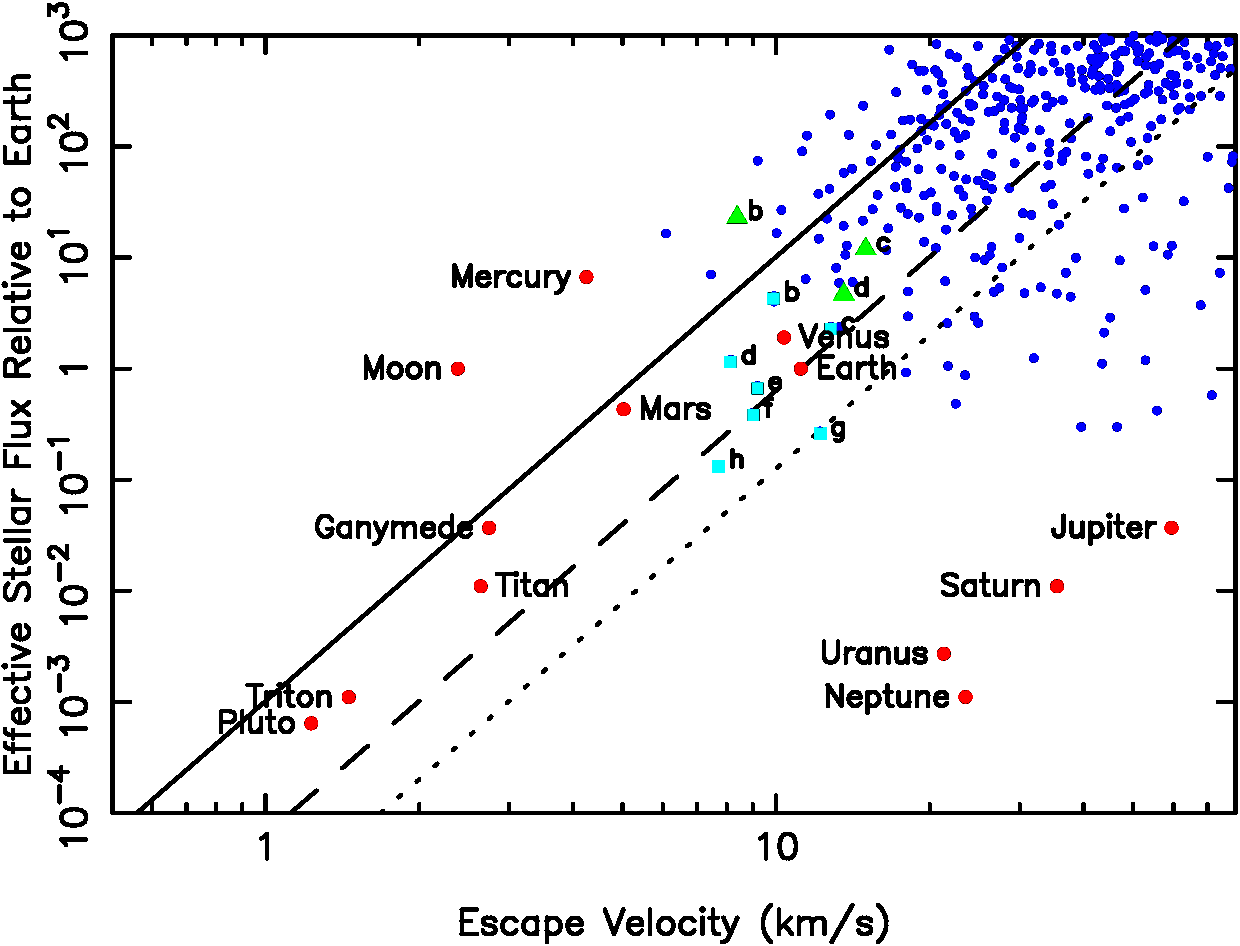}}
  \caption{Representation of the cosmic shoreline \citep{zahnle2017}, shown as a solid line. Additional dashed and dotted lines assume XUV flux 16 and 256 times higher than solar, respectively. Solar system bodies are shown as red circles, and exoplanets as small blue dots. The L~98-59 and TRAPPIST-1 planets are indicated by green triangle and cyan squares, respectively.}
  \label{fig:cosmic}
\end{figure}

Furthermore, the relative size and insolation fluxes of the L~98-59 planets places them within an interesting regime of potentially significant atmospheric loss. Shown in Figure~\ref{fig:cosmic} is a representation of the cosmic shoreline, which explores the relationship between XUV flux and the escape velocity for Solar System objects with respect to atmospheric retention \citep{zahnle2017}. The Solar System objects are marked as red circles and the solid line is the cosmic shoreline, based on the empirically determined estimates of \citet{zahnle2017}. Objects to the left of the solid line tend to lack atmospheres, whereas objects to the right do have atmospheres, albeit tenuous for those close to the line. We include two additional lines, dashed and dotted, that represent XUV factors 16 and 256 higher than solar, respectively \citep{roettenbacher2017}. We also include known exoplanets, shown as small blue dots, for which the required data are available from the NASA Exoplanet Archive \citep{akeson2013}. The location of the TRAPPIST-1 and L~98-59 planets are indicated by cyan squares and green triangles, respectively. Although the TRAPPIST-1 planets have a relatively high risk of significant atmospheric loss \citep{roettenbacher2017,cohen2018}, the L~98-59 planets also fall within a flux regime that means they are interesting case studies in assessing the atmospheric evolution for terrestrial planets in the presence of a relatively high XUV environment. \\

Herein, we explore the potential for atmospheric detection and characterization across a range of varying scenarios for L 98-59 b, c, and d. The feasibility of atmospheric characterization is determined first by selecting the most favorable instruments to conduct transit spectroscopy measurements of each planet. Theoretical observations of the planets are conducted using HST to determine what we could detect in the near-term, while simulated JWST observations look into the possibilities of the intermediate future. The atmospheres considered here do not represent the full spectrum of potential outcomes, but are merely a representation of atmospheres which have been motivated by their ability to produce measurable spectra, given the known planetary parameters. \\

The structure of the paper is as follows: Section \ref{sec:methods} provides a detailed overview of the methods and inputs considered in the simulation of spectra using HST and JWST. Section \ref{sec:hst} considers the prospective atmospheric detection that can be carried out with HST, while Section \ref{sec:jwst} explores the atmospheric signatures that will be targets of interest for near-future characterization with JWST. A discussion of the results can be found in Section \ref{sec:discussion}, with conclusions following in Section \ref{sec:conclusion}.

\section{Methods} 
\label{sec:methods}
In the following subsections, we describe our methods for assessing the detectability and signal-to-noise ratio (S/N) of various spectral  signatures for the L 98-59 system with numerous instruments and observational modes available to HST and JWST. The elemental and molecular composition of each planetary environment is simulated and analyzed using the Planetary Spectrum Generator (PSG), an online\footnote{PSG is available at \url{https://psg.gsfc.nasa.gov/}} radiative-transfer suite that computes synthetic transit spectra for a wide range of planetary atmospheres \citep{Villanueva2018}. We aim to determine how feasible the detection and characterization of atmospheres dominated by H$_2$ or H$_2$O would be for HST, and we assess how the detection may be improved with observations via JWST. Simulated JWST observations also include scenarios where the atmospheres are dominated by CO$_2$ or O$_2$, representing post-runaway and/or desiccated states. In some cases, we also include the presence of aerosols. We consider an atmosphere to be detected when sufficient S/N is achieved on the strongest molecular feature in the spectrum at a 5$\sigma$ confidence level. All simulations include a panel that details the relative contribution of each molecule to the overall spectrum of the atmosphere. These spectra are calculated by subtracting the spectrum without the molecule from the full spectrum.

\subsection{Simulating Transit Spectra with PSG}
Our transit simulated spectra with PSG integrate molecular and aerosols extinctions via accurate spectroscopic methods and parameters, which include a realistic treatment of the radiative-transfer in layer-by-layer pseudo-spherical geometry. For this investigation, the molecular spectroscopy is based on the latest HITRAN database \citep{gordon2017} integrated via the correlated-k method, which is complemented by UV/optical data from the Max-Planck Institute of Chemistry database \citep{kellerrudek2013}. In addition to the CIA bands available within the HITRAN database, the MT\_CKD water continuum is characterized as H$_2$O-H$_2$O and H$_2$O-N$_2$ CIAs \citep{Kofman2021}. PSG also includes a large database of scattering and absorptive properties from many known aerosols -- both measured $in$ $situ$ on Earth as well as laboratory measured values for other solar system materials -- which can be added ad-hoc to any synthetic transit spectrum. Further details on the treatment of aerosols for the simulations presented here can be found in Section~\ref{subsub:aerosols}. In generating each spectrum, we input basic parameters of the star and each planet, summarized in Table~\ref{tab:starandplanet}. Planetary atmospheres are designed by adjusting a range of parameters fully described in Section~\ref{sub:atmosphere}.

\subsection{Stellar and Planetary Parameters}
\label{sub:starplanetparams}
In all simulations, planetary parameters are derived from the system's discovery \citep{Kostov2019}, as well as \cite{cloutier2019}, whose work constrained the masses for each planet. Current RV measurements only provide an upper limit for the mass of L 98-59 b of $<$1.01 M$_\earth$, here we assume 0.45 M$_\earth$, or 2.70x10$^{24}$ kg for an Earth-like bulk composition \citep{Lopez_2014}. For planetary studies, the planetary equilibrium temperature is estimated assuming either an Earth-like Bond albedo of $A$ =  0.3 or an unrealistic $A$ =  0.0   \citep[e.g.][]{Borucki2012,Anglada2016,Ditmann2017,Gillon2017}. For planets orbiting an M dwarf, the Bond albedo would be lower than the equivalent planet placed around a Sun-like star. This is because of the IR shift of the incident spectrum, which would be less efficiently scattered or reflected by most plausible planetary atmospheres. Here, we use the Earth Bond albedo value of $A$ =  0.3, keeping in mind that this is a nominal estimate. A summary of the stellar and planetary system parameters as model inputs for each simulation is shown in Table \ref{tab:starandplanet}. 

\begin{table*}[ht]
\centering 
\caption{Stellar and planetary system parameters modeled} 
 \label{tab:starandplanet}
\begin{tabular}{c c c  c }
\hline
\hline
Stellar Parameter & Modeled \\

\hline
Spectral type & M3V \\   
Mass (M$_\sun$) & ${0.312\pm0.031}$ \\ 
Radius (R$_\sun$) & ${0.314\pm0.014}$  \\
Effective temperature (K) &  ${3412\pm49}$  \\
$J$ (mag) & ${7.933\pm0.027}$ \\ 
\hline
\hline 
Planetary Parameter & L 98-59 b & L 98-59 c & L 98-59 d  \\
\hline 
Semi-major axis (AU) & 0.02282 & 0.0317 & 0.0506  \\
Mass (M$_\earth$) & 0.45 & $2.42\ \pm  0.35$ & $2.31\ \pm  0.46$ \\ 
Radius (R$_\earth$) & $0.80\ \pm  0.05$ & $1.35\ \pm  0.07$ & $1.57\ \pm  0.14$ \\ 
Orbital period (days) & 2.2531 & 3.6904 & 7.4512  \\
Transit duration (hours) & 1.02 & 1.24 & 0.91   \\
Inclination (deg) & 88.7 & 89.3 & 88.5 \\ 
Equilibrium temperature (A = 0.3) (K) & 558 & 473 & 374 \\ 
Insolation & 23.9 & 12.4 & 4.85 \\ 
Mean density (g/cm$^3$) & 4.9 & 5.4 & 3.3 \\
Surface gravity (m/s$^2$) & 6.9 & 13.1 & 9.2 \\
\hline

\hline
\end{tabular}
\end{table*} 
 
 \

\subsection{Atmospheric Parameters} 
\label{sub:atmosphere}
This work considers four potential planetary environments, each varying in elemental composition, while some expand to include additional caveats such as the presence of clouds and hazes. For all three planets, we simulate clear and cloudy steam atmospheres, desiccated atmospheres composed of O$_2$/O$_3$, and clear and cloudy Venus-like atmospheres dominated by CO$_2$. An additional clear low mean-molecular weight atmosphere is also simulated for L 98-59 c and d. \\ 

The structure of the atmosphere is described in PSG layer-by-layer, with information for each layer including pressure (bar), temperature (K), molecular abundances (molecules/molecules), and aerosols abundances (kg/kg) with respect to the total gas content. Altitudes are computed employing the hydrostatic equation. Computation of layer-by-layer integrated column densities (molecules/m$^2$) and aerosols mass densities (kg/m$^2$) are then computed along the transit slant paths employing a pseudo-spherical geometry. With no constraints on planetary surface temperatures, some simulations use an isothermal temperature profile set at the planetary equilibrium temperature. We note that this method is conservative as it does not account for possible additional heat sources or temperature inversion and results in a possible under-evaluation of the atmospheric scale height. \\ 

Each atmosphere is configured with varying inputs:  

\begin{itemize}
    \item \textbf{H$_2$-Dominated:} The pressure/temperature (P/T) profiles are calculated based on a non-gray analytical model \citep{Parmentier_2014} according to the planet's equilibrium temperature and gravity. Molecular abundances are then derived layer-by-layer by considering the equation-of-states (EOS) and chemical equilibrium computed by \cite{Kempton_2017}, which does not include disequilibrium chemistry, nor photochemistry. The resulting atmosphere is detailed in Table \ref{tab:atmospheres}, and the concentrations of H$_2$O and CH$_4$ remain nearly constant (within 20 ppm) over altitude. Each simulation is representative of a cloud-free atmosphere.
    \item \textbf{Steam Atmosphere:} For a water-rich steam atmosphere, we assume purely radiative, isothermal atmospheric cases, with and without the presence of aerosols. Previous studies have used this isothermal approximation only for the mesosphere and above \citep{Kasting1988, Kopparapu2013, Leconte2013a, Marcq2017,Turbet2019}. Even if the surface temperature is extremely hot, the top of the atmosphere temperature is always fairly cold. Since we are not using a climate model to predict the tropospheric temperature, we have extended this approximation to the troposphere as well. This should not have an impact on the transmission spectra because in a hot and moist atmosphere, the lowest layers are opaque to infrared radiation and could therefore not be probed using transmission spectroscopy. The volume mixing ratios of H$_2$O and N$_2$ are kept constant, whose values are shown in Table \ref{tab:atmospheres}. In simulations where aerosols are included, their treatment is summarized in subsection \ref{subsub:aerosols}. We also note that the temperature of the isotherm in the mesosphere and above would likely be higher than the planet's equilibrium temperature that we have assumed in this work. Thus, we may have underestimated the scale height and the atmospheric transit depth for steam atmospheres.
    \item \textbf{Post-runaway Greenhouse:} A Venus-like atmosphere with a provided vertical profile dominated by CO$_2$ is simulated from the use of a preloaded template in PSG, whose parameters are derived from \cite{Ehrenreich2012,Vandaele2020,Bierson2020}. To approximate the temperature of the tropopause, we set the top-of-atmosphere (TOA) to each planet's skin temperature ($T_{s}=T_{e q}\left(\frac{1}{2}\right)^{1 / 4}$; b = 469K, c = 398K, d = 315K). In the clear case, the volcanic aerosols included in the cloudy case from \cite{palmer1975} are removed. For simulations where aerosols are included, their treatment is summarized in subsection \ref{subsub:aerosols}.
    \item \textbf{O$_2$-Desiccated:} For a post-runaway desiccated planet, we assume an isothermal atmosphere that is rich in abiotic O$_2$, where we expect to see O$_3$ formation from the photochemical processing of this O$_2$. To properly account for the amount of O$_3$ present in the atmosphere, we use the photochemical module of Atmos, a coupled 1D photochemical-climate model \citep{Arney2016}. Its photochemical module is based on a code originally developed by \cite{Kasting1979} and has been significantly modernized as described in \citet{Zahnle2006}. 
    The photochemical model has been additionally updated as described in \citet{Lincowski2018}, where it has also been validated on Earth and Venus. We assume an atmosphere dominated by O$_{2}$ with trace amounts of O$_{3}$ (calculated by the model) and no other gases. We consider total pressures of 1 and 10 bars. Because no complete calibrated stellar spectrum of L 98-59 is available, for the purposes of modeling the photochemistry of these O$_{2}$-rich atmospheres only, we substitute as the input stellar spectrum the panchromatic spectrum of GJ 581 (also M3V) from the MUSCLES Survey (v21) \citep{France2016, Youngblood2016, Loyd2016} scaled to the fluxed given in Table \ref{tab:starandplanet}. The output chemical profiles from Atmos are then provided to PSG to construct a simulated spectrum. 
\end{itemize}
A summary of the atmospheric parameters as model inputs for each simulation is shown in Table \ref{tab:atmospheres}. The mean molecular weight (MMW) of each gas is given in g/mol. 

\begin{table*}[ht]
    \centering 
    \caption{Atmospheric parameters modeled}
    \begin{tabular}{c c c c c}
    \hline
    \hline
    Atmospheric state & Aerosols & Gases & MMW (g/mol) & Surface pressure (bar) \\

    \hline
    H$_2$-dominated & none & 83.7\% H$_2$, 16.2\% He, trace gases$^{1,2}$ & 2.36 & 1  \\
    Steam, clear sky & none & 90\% N$_2$, 10\% H$_2$O & 18.02 & 1  \\
    Steam, hazy & Titan tholins & 90\% N$_2$, 10\% H$_2$O, trace organic haze & 18.02 & 1  \\
    CO$_2$-dominated, clear sky & none & 96.5\% CO$_2$, 3.6\% N$_2$, trace gases$^3$ & 44 & 92  \\ 
    CO$_2$-dominated, cloudy & H$_2$SO$_4$ clouds & 96.5\% CO$_2$, 3.6\% N$_2$, trace gases$^4$ & 44 & 92  \\
    O$_2$-desiccated & none & 99.9\% O$_2$, trace O$_3$$^5$ & 32 & 1, 10 \\
    \hline
    \hline
    \end{tabular}
\begin{tablenotes}
    \item[1] $^1$L 98-59 c: 663.6 ppm H$_2$O, 488.3 ppm CH$_4$, trace CO, CO$_2$, O$_2$
    \item[2] $^2$L 98-59 d: 692.7 ppm H$_2$O, 488.5 ppm CH$_4$, trace CO, CO$_2$, O$_2$
    \item[3] $^3$trace CO, H$_2$O, SO$_2$, O$_2$, O$_3$
    \item[4] $^4$trace CO, H$_2$O, SO$_2$, O$_2$, O$_3$, H$_2$SO$_4$
    \item[5] $^5$see Figure \ref{fig:o3-concentration}
    \end{tablenotes}
    \label{tab:atmospheres}
    \end{table*}

\subsection{Treatment of Clouds/Hazes} 
\label{subsub:aerosols}
Both clouds and hazes are ubiquitous throughout the solar system, while their presence on various exoplanets is widely inferred from observations \citep[e.g.,][]{kreidberg2014}. However, the treatment of clouds and hazes in atmospheric models for transit spectra simulations varies greatly, from fully self-consistent photochemical models \citep[e.g.,][]{morley2015,Lincowski2018,Meadows2018a} to highly parameterized scattering functions \citep[e.g.,][]{sing2016}. By integrating the layer-by-layer aerosols abundances (kg/kg) and spectroscopic scattering models, PSG solves for the radiative transfer across the line-of-sight. For aerosols employing the HRI \citep[\text{HITRAN Refractory Index};][]{Massie2013} database, PSG utilizes pre-computed Mie scattering models for different particle sizes in the form Henyey-Greenstein coefficients \citep{Henyey1941}, which are internally expanded into Legendre coefficients to be ingested by the radiative-transfer suite. Laboratory results for the optical properties of exoplanet hazes do not yet exist, so in this work, we include clouds and hazes in our models which have either direct solar system analogues or from which we approximate based off solar system analogues together with existing laboratory results on particle size, chemistry, and production rate for exoplanet hazes. \\

For CO$_2$ atmospheres, we use the refractive indices of  \citet{palmer1975} for sulfuric acid at 75\% concentration for Venus-like clouds in PSG, placing clouds between 1 and 0.01 bar, at 1 kg/m$^{3}$ mass density, with particle sizes of 1 $\mu$m. These values follow from Figure 1 of \citet{Lebonnois2017}, which provides a model of the vertical structure of Venus' atmosphere. Here, we choose to use Venus clouds as a possible representative case for lack of other constraints. Venus-like clouds are a potentially pessimistic case, as the high irradiance of the L 98-59 planets may cause true Venus clouds to burn off and dissipate more readily than in the upper atmosphere of Venus itself. However, recent laboratory results regarding haze formation in warm CO$_2$ atmospheres suggests high sulfuric acid production, which could give rise to clouds \citep{Vuitton2020}.  We also note that the formation of sulfuric acid clouds requires the presence of trace amounts of H, which may have escaped from this atmosphere. Nevertheless, we use such clouds in our cloudy CO$_2$ PSG simulations to explore the effect reasonable clouds might have on observations made by JWST. 

\begin{figure*}[t]
    \centering
    \gridline{\fig{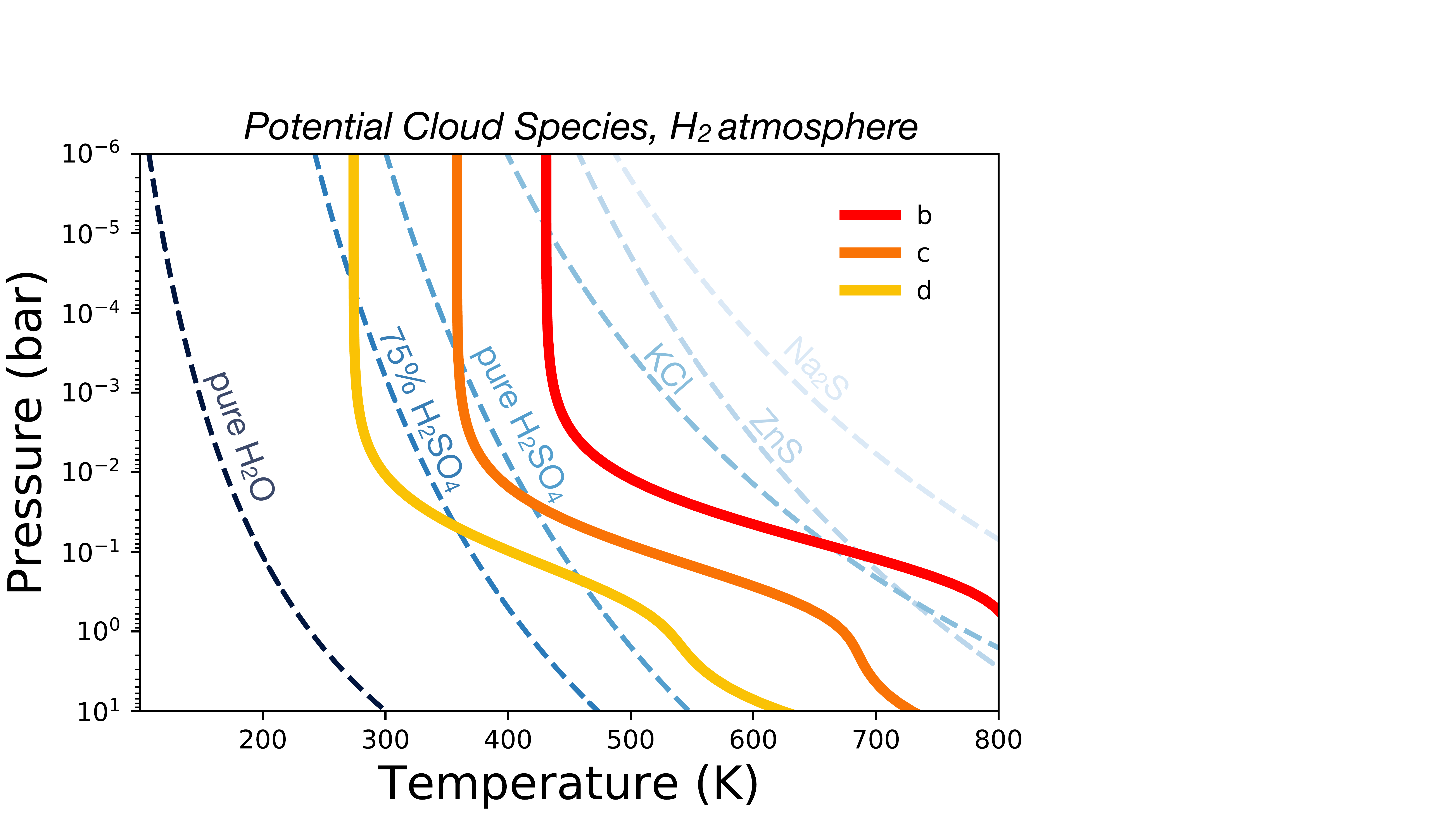}{0.49\linewidth}{}
          \fig{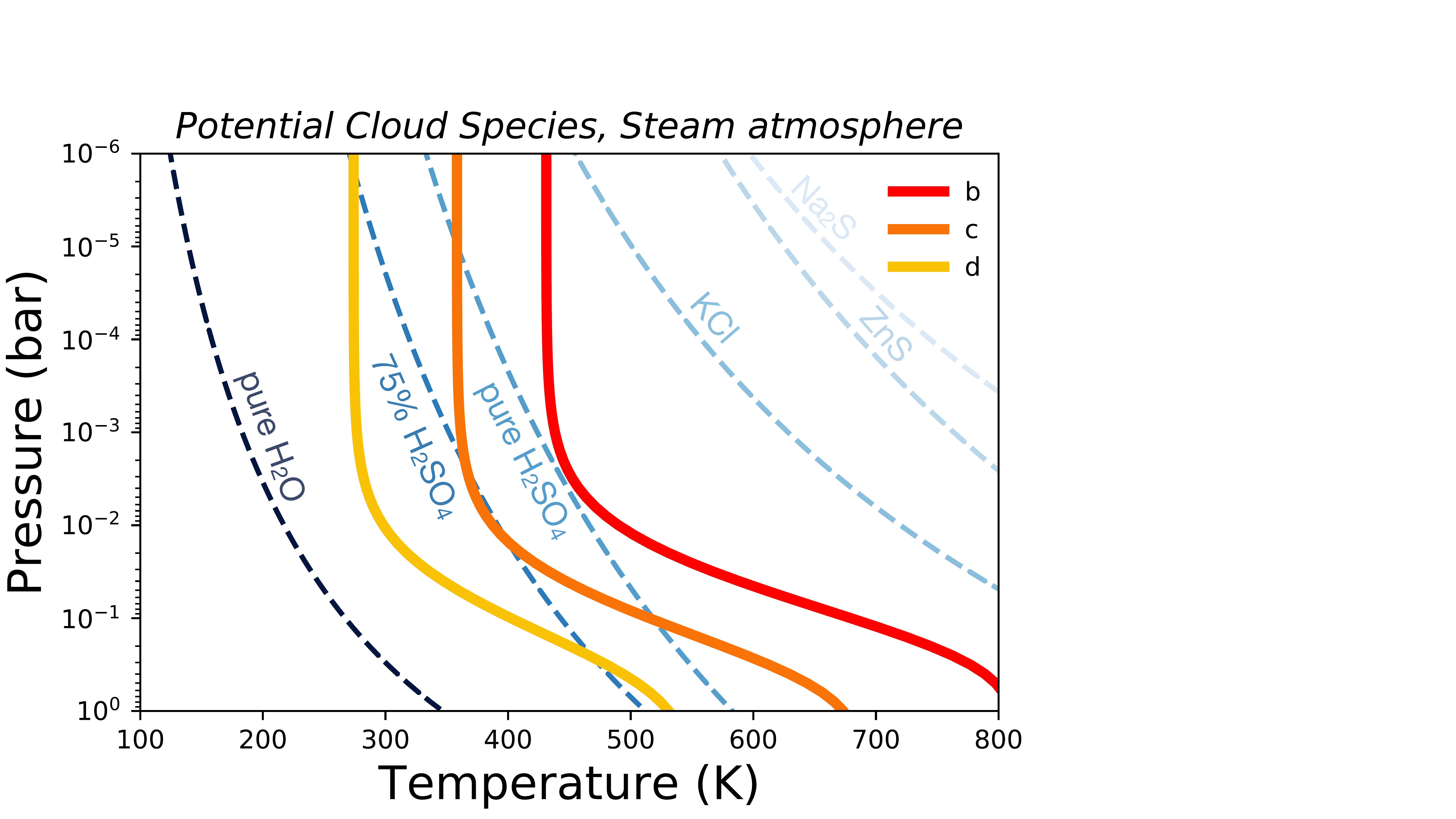}{0.49\linewidth}{}}
    \caption{Temperature pressure profile models of each planet in the L98-59 system compared to theoretical cloud vapor pressure curves. Where a planet profile crosses a vapor pressure curve, cloud condensation is expected. (Left) solar metallicity profiles and saturation curves; (right) water-rich profiles and cloud curves.}
    \label{fig:clouds}
\end{figure*}

We show in Figure \ref{fig:clouds} the saturation vapor pressure profiles (derived following \cite{morley2013}) of pure water, pure H$_2$SO$_4$, a 75\% sulfuric acid concentration, KCl, ZnS, and Na$_2$S together with parameterized temperature-pressure profiles of the three L 98-59 planets. Of all known potential cloud species, only some sulfuric acid cloud (with some level of water concentration) is likely to form in the atmospheres of these planets, which is why we include them in our pessimistic cloudy case for a CO$_2$ atmosphere. We use these models as a guide for which cloud species to consider rather than as a self-consistent framework upon which to model clouds. If a saturation vapor pressure curve is crossed, the potential for such clouds exists, but we do not prescribe where they must be in the atmosphere. Even if clouds form lower in the atmosphere than observations probe, atmospheric mixing may loft them to higher altitudes where they may impact observations \citep{morley2013}. Saturation vapor pressure profiles are metallicity-dependent; therefore, ZnS or KCl clouds could be possible for L 98-59 b in a solar metallicity atmosphere (Fig \ref{fig:clouds} (left)). However, L 98-59 b is unlikely to retain such an  H$_2$-rich atmosphere (Section \ref{subsec:h2hst}), and the temperature-pressure profile of a high-metallicity, H$_2$O- or CO$_2$-rich L 98-59 b does not cross the curves of these cloud species at all (Fig \ref{fig:clouds} (right)). Given these considerations, we do not consider KCl or ZnS cloud species in our models.   \\ 

For steam atmospheres, laboratory experiments have shown substantial haze production \citep{Horst18}, and the compositions of these hazes appear to be complex organics \citep{Moran2020}. For hazes in steam atmospheres, we use the refractive indices of \citet{Khare1984} for Titan-based organic haze which is pre-loaded in PSG from the HRI database. The atmospheric parameters for Titan are derived from \cite{Teanby2006}. \citet{Khare1984} measurements are one very specific outcome of haze formation and composition produced via a room-temperature 90\% N$_2$, 10\% CH$_4$ experiment done under ambient Earth atmosphere. Therefore, it should be treated with heavy skepticism as a basis for extrapolation to exoplanet atmospheric hazes. However, as there are currently no laboratory measurements of optical properties of any exoplanet hazes relevant to the bulk atmospheric compositions explored here, the use of \citet{Khare1984} was chosen as a necessary proxy, with the caveat that further measurements and models should be performed as additional theoretical data become available. Laboratory measurements specific to exoplanets like L 98-59 will be required to make true interpretations of atmospheric observations, rather than the exploration of generally reasonable but unconstrained scenarios as we have done in this work. We do not explore condensation clouds in steam atmospheres because of the temperature regimes encompassed by the atmospheres of the L 98-59 planets. The planetary atmospheres, even high in the atmosphere, never get cool enough for water clouds to condense, and other cloud species \citep[\text{i.e., those shown by other works such as}][]{morley2013} form only at hotter temperatures in a higher metallicity steam atmosphere as we show in Figure \ref{fig:clouds}. \\

We choose not to include any aerosols in either our H$_2$ atmospheres or our desiccated O$_2$ atmospheres. Laboratory results have shown that hydrogen atmospheres have poor haze production efficiency in comparison to atmospheres dominated by H$_2$O or CO$_2$ \citep{Horst18}, and previous modeling work shows that the amount of haze needed in hydrogen atmospheres to significantly impact observations of terrestrial planets is unrealistic \citep{Moran2018}. While previous studies considered KCl or ZnS clouds in H$_2$ atmospheres \citep{morley2013}, more recent work shows that only silicate clouds are likely to substantially impact the spectra of H$_2$ atmospheres due to nucleation energy barriers for salt cloud formation \citep{Gao2020}, in addition to this only being relevant for L 98-59 b. As none of the L 98-59 planets are hot enough for silicate cloud formation, we ignore clouds for our H$_2$ atmospheres. Although further observations are required to evaluate whether haze formation is efficient or inefficient in H$_2$ atmospheres, the detection or absence of N$_2$-N$_2$ or N$_2$-X CIA features may help distinguish between types of atmospheres that are either hazy H$_2$ dominated or N$_2$-H$_2$O dominated \citep{Schwieterman2015}. For O$_2$ atmospheres, we do not include aerosols because we find no photochemical modeling results, laboratory results, or solar system analogues upon which to base a possible aerosol layer in a desiccated O$_2$ atmosphere. However, this may be an interesting avenue to explore in future work. 

\subsection{Instrumental Parameters and Sensitivity Analysis}
Once the atmospheric framework for the planet of interest is selected, PSG can be used to model the instrument parameters for any given observatory. Each telescope and instrument considered in this work uses the input parameters defined by the database of preloaded models within PSG. PSG accounts for system throughput and presents the user with a table that describes these values, all of which are found in the instrument and detector parameters. This work uses PSG to model transmission spectroscopy observations specifically using multiple instruments from HST and JWST. Although fringing and other systematic sources of noise have been identified in these observational modes, their overall effect on the actual reduced data are not yet sufficiently quantified, and these effects were not included in our noise simulator and could lead to an overestimation of the S/N (see Section \ref{sub:noise}). We have benchmarked PSG's noise simulator by validating our JWST models against PandExo \citep{Batalha2017} where we find PSG and PandExo to be within 10\% agreement for all JWST instruments. \\

Instrumental inputs include the following: 

\begin{itemize}
    \item \textbf{HST/WFC3:} Simulated observations of transmission spectroscopy with HST are conducted using the near-infrared ($1.1-1.7\ \mu$m) G141 grism on the Wide-Field Camera 3 (WFC3) instrument (R = 130). 
    \item\textbf{JWST/NIRISS SOSS:} Modeled simulations of exoplanet atmospheres show that observations with the Near-Infrared Imager and Slitless Spectrograph (NIRISS) operating in Single Object Slitless Spectroscopy (SOSS; R = 700) are required to cover the $0.6-2.8\ \mu$m wavelength range for targets that are too bright to be observed with NIRSpec Prism, which would be saturated in one read of the detector when observing a star as bright as L 98-59 (K = 7.1) \citep{BatalhaLine2017}.
    \item\textbf{JWST/NIRSpec G395H:} To simulate spectral features within the $2.87-5.14\ \mu$m wavelength range, we use NIRSpec with the G395H disperser (R = 2700), as an observation with both NIRISS and NIRSpec G395H yields the highest content spectra with the tightest constraints \citep{BatalhaLine2017}.
    \item \textbf{JWST/MIRI LRS:} Beyond the extent of NIRISS/NIRSpec's wavelength range, certain gaseous features such as the O$_2$-O$_2$ CIA feature at $6.4\ \mu$m are only accessible through the use of the Mid-InfraRed Instrument \citep[\text{MIRI};][]{Bouchet2015} low-resolution spectrometer \citep[\text{LRS; R = 100};][]{Kendrew2015}. MIRI LRS has both a camera and a spectrograph that perform between the wavelength range of $5.0-28.0\ \mu$m, with transit observations ending at $12.0\ \mu$m.
\end{itemize}

\ 

For the sensitivity analysis, PSG includes a noise calculator that accounts for the noise introduced by the star or source itself (N$_{source}$), and the background noise (N$_{back}$, e.g., zodiacal dust) following a Poisson distribution with fluctuations depending on $\sqrt{N}$ with $N$ the mean number of photons received, the noise of the detector (N$_{D}$) and the noise introduced by the telescope (N$_{optics}$). The total noise is then defined as  $N_{total}=\sqrt{N_{source}+ 2(N_{back}+N_D+N_{optics})}$.  \\ 

In all synthetic spectra shown within this work, the resolving power (R = $\lambda/\Delta\lambda$) selected for each HST and JWST instrument is varied to accommodate the most efficient detection of the strongest molecular feature found within the spectrum of a given atmosphere, based on Table 4 in \cite{Wunderlich2019}. The chosen R value for a specific detection is noted in the results (Tables \ref{tab:hstresults} and \ref{tab:jwstresults}). To calculate the S/N and the number of transits needed for a 5$\sigma$ detection, the resolving power is optimized by adjusting the binning of the strongest feature in each simulation to maximize its S/N. \\

S/N (considered "SNR" in the following equations) is calculated by subtracting the nearest continuum value from the highest value of any given spectral band. This value, considered the "relative depth" of a feature, therefore differs between the visible (VIS; dominated by the Rayleigh slope), near infrared (NIR) and mid-infrared (MIR). The relative depth can then be divided by the noise value at the given band to determine SNR. For a source dominated noise limit, the number of transits needed to achieve a X$\sigma$ detection is computed as
\begin{equation}\label{eq:SNRi}
    N_{transits}^{X\sigma}= N_i*(X/SNR_i)^2,
\end{equation}
with $X\sigma$ the confidence level of value $X$, N$_i$ is the initial number of transits at which the SNR$_i$ is computed. If the SNR$_i$ is estimated from 1 transit, the $N_i=1$ and the Eq. \ref{eq:SNRi} can be simplified as
\begin{equation}
    N_{transits}^{X\sigma}= (X/SNR_i)^2.
\end{equation} \

\section{Prospects and Results for Characterization with HST} \label{sec:hst}
Thus far, HST atmospheric follow-up of small planets has focused on detecting the broad water peak at 1.4 $\micron$ with HST's Wide Field Camera 3 IR grisms (e.g., GJ~436b \citep{knutson436}, HD~97658b \citep{knutson976}, GJ~1214b \citep{kreidberg2014}, GJ~1132b \citep{southworth2017}, TRAPPIST-1 \citep{deWit2016,deWit2018}, GJ 3470 b \citep{Benneke2019a}, K2-18b \citep{Benneke2019b,Tsiaras2019}, HD 106315 c \citep{Kreidberg2020}, HD 3167 c \citep{MikalEvans2021}, and LHS 1140 b \citep{Edwards2021}). However, this method of exploring whether or not such worlds have atmospheres, or further, what the chemical constituents of these atmospheres are, has been plagued by weak, muted molecular features due to either high mean molecular weight atmospheres or high cloud decks \citep[e.g.,][]{sing2016,wakeford2019rnaas}. Although planets with detected hydrogen envelopes are slightly larger ($>$ 2 R$_\earth$) than the planets in the L 98-59 system, transit observations with HST would provide a first look at determining whether or not these planets contain atmospheres. With current ongoing efforts through HST's general observer (GO) program, we soon expect to place these initial constraints on L 98-59 b--d (HST-GO-15856 and HST-GO-16448, PI Barclay) and specifically constrain abundances if they are dominated by abundances of H$_2$ or H$_2$O. \\ 

In the following subsections, we present simulated transmission spectra of each planet via WFC3. Details on S/N and transit values for individual spectral features are found in Table \ref{tab:hstresults}. 

\subsection{Detecting a Low Mean-Molecular Weight Atmosphere with HST}
\label{subsec:h2hst}
Hitherto, atmospheric characterization of exoplanets has revealed an abundance of close-in planets with atmospheres holding substantial amounts of hydrogen/helium. While small planets are unlikely to retain primordial hydrogen-rich atmospheres \citep{Rogers2015, Fulton2017}, secondary atmospheres could form from volcanic outgassing or delivery of volatiles from comets. The presence of exoplanets with volatile atmospheres at short orbital periods raises the question as to whether these atmospheres are stable \citep{Koskinen2007}. For hydrogen-dominated atmospheres, the high UV flux close to the star dissociates the molecular hydrogen, resulting in atomic hydrogen dominating the upper regions \citep{Yelle2004,Garcia2007}. In these atomic regions, heating typically results in gas temperatures on the order of 5,000--10,000 K. At such high temperatures, the upper atmospheres of close-in exoplanets are weakly bound, and the gas becomes more susceptible to escaping the planet's gravity. In many cases, and especially for small, low-mass planets such as those within the L 98-59 system, the total time-integrated high-energy flux (high-energy exposure) a planet receives over its lifetime of billions of years is a significant fraction of its gravitational binding energy \citep{Lecavalier2007}. This available energy means that unlike solar system planets, where atmospheric escape is vital for shaping the chemical evolution of a planet's atmosphere \citep{Lammer2008}, atmospheric escape can affect the evolution of a planet's bulk composition for close-in exoplanets. Modeling work suggests that this atmospheric escape is a strong driver in the evolution of many close-in planets \citep{Owen2019}. \\ 

To determine the plausibility of each planet holding an atmosphere dominated by hydrogen, we set the root mean square velocity of H$_2$ equal to the escape velocity of the planet. The following formula is used to calculate the escape temperature of H$_2$ for each given planet, based on its size:
\begin{equation}\label{Tesc}
    T_{\mathrm{esc}}>\frac{1}{54}\frac{G M_{p}m}{kR_{p}}
\end{equation}
Due to its low expected gravity, an H$_2$ atmosphere on L~98-59~b would be highly vulnerable to atmospheric escape. When considering a bond albedo = 0.3, a calculation of the escape temperature of hydrogen on L~98-59~b shows that the equilibrium temperature (558~K) is higher than the escape temperature of H$_{2}$ molecules (342~K) and thus it is not likely that the planet retains such an atmosphere. For this reason, an H$_2$ dominated atmosphere on L 98-59 b is not investigated within this work. However, the equilibrium temperatures of L 98-59 c and d (473~K; 374~K, respectively) allow for retention of secondary H$_2$ atmospheres, given that their equilibrium temperatures are lower than their escape temperatures for H$_2$ (509~K; 402~K, respectively). \\ 

\begin{figure*}[t]
    \centering
    \resizebox{13cm}{!}{\includegraphics{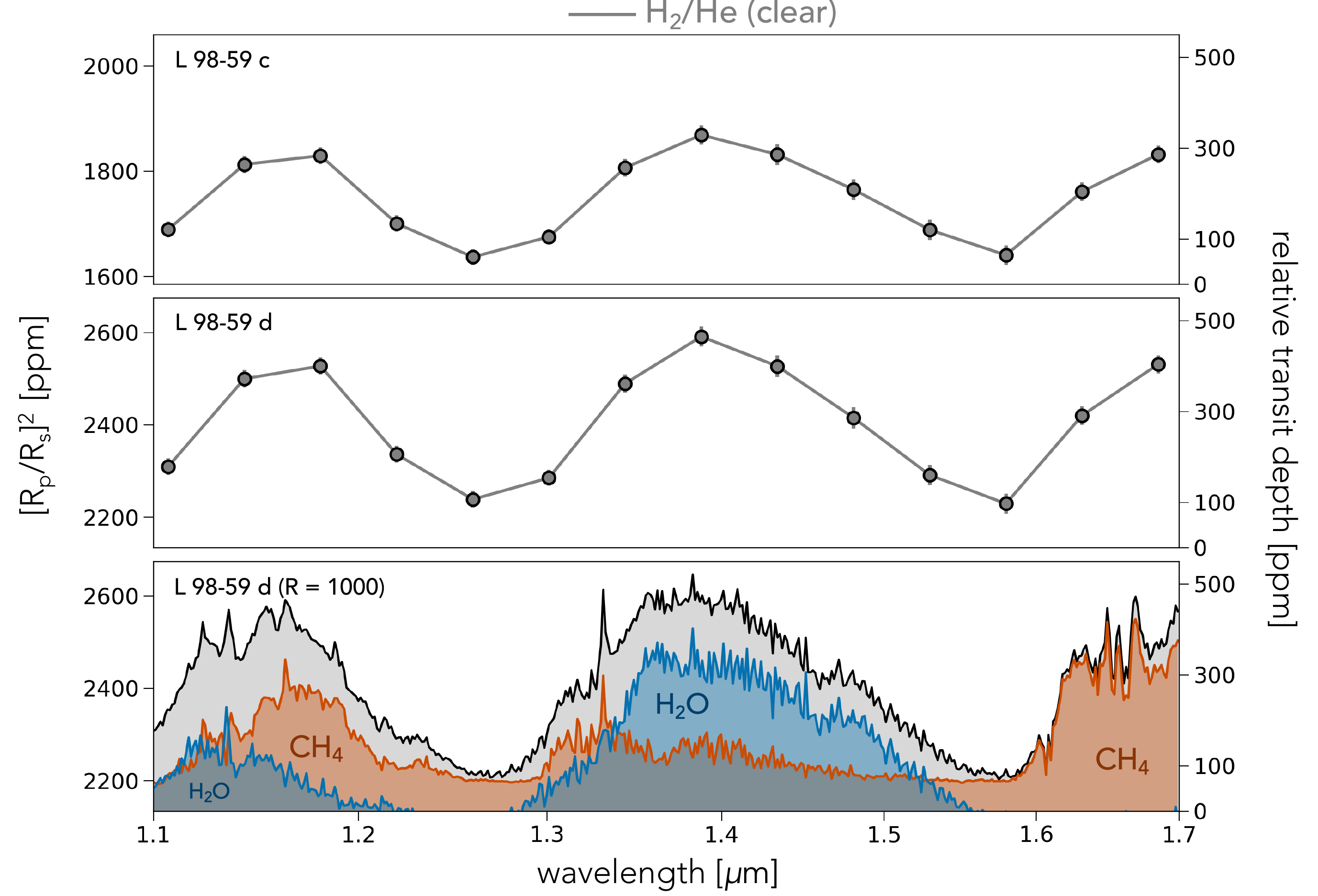}}
    \caption{One transit of a H$_2$/He-dominated atmosphere simulated with HST/WFC is shown for L 98-59 c (top) and L 98-59 d (middle) in grey. Error bars are shown for observations simulated at R = 30. The gaseous species corresponding to the molecular features in each planetary spectrum are shown (bottom) at a higher (R = 1000) resolution. Here, the black line represents the corresponding simulated spectrum for L 98-59 d (middle), while the grey shading indicates the sum of all gases present. The absorption of H$_2$O is shown in blue and CH$_4$ in orange.}
    \label{fig:h2dominated_atmosphere_spectrum}
\end{figure*}

Figure \ref{fig:h2dominated_atmosphere_spectrum} shows the simulated spectrum for one transit of a clear atmosphere dominated by H$_2$/He for L 98-59 c and L 98-59 d. The bottom panel accompanies these spectra to show the detailed molecular composition of each planetary atmosphere, using L 98-59 d as an example. Although the atmospheres contain a multitude of trace gases (Table \ref{tab:atmospheres}), only H$_2$O and CH$_4$ are shown individually as they are primarily responsible for creating the absorption features visible in each modeled spectrum. 
 
\

On L 98-59 c and d, a clear-sky atmosphere dominated by H$_2$/He has multiple spectral features that can be detected and characterized with WFC3. Although the features found in the $1.1 - 1.3\ \mu$m range are blended by H$_2$O and CH$_4$, the peaks at 1.38 and $1.68\ \mu$m can be attributed to the absorption of H$_2$O and CH$_4$, respectively. These peaks provide ample signal in both planets, notably for L 98-59 d, whose lower density increases S/N and scale height substantially. In a single transit, HST could potentially rule out a featureless transmission spectrum on such an atmosphere for both planets with a 5$\sigma$ confidence level and S/N $\geq$ 10, providing ample information for a minimal allocation of time. Moreover, a physically realistic amount of clouds and/or hazes in a H$_2$ atmosphere are anticipated not to mute features beyond HST's capabilities \citep{Moran2018}.

\subsection{{\texorpdfstring{Detecting H\textsubscript{2}}OO with HST}}
Beyond the ability to detect the presence of an atmosphere or rule out the possibility of one dominated by H$_2$/He, HST observations may be used to detect H$_2$O absorption in a steam atmosphere. It is of particular interest to investigate this possibility on L 98-59 b, where the mean-molecular weight of a steam atmosphere makes it much more resilient to atmospheric escape in comparison to one dominated by H$_2$. While H$_2$-dominated sub-Neptunes with equilibrium temperatures similar to L 98-59 b (${\rm T_{eq}\sim558\ K}$) are sometimes cloudy/hazy (e.g., as in the case of GJ~1214~b; \citealt{kreidberg2014}), a steam atmosphere on the terrestrial L 98-59 b would likely be very different. An outgassed atmosphere from a volatile-rich surface would have few trace species (e.g., Na$_2$S, KCl) that condense at these temperatures to form clouds (and the atmosphere is likely too hot for water itself to condense) \citep{morley2013}. While lab experiments for colder steam atmospheres show substantial haze production rates \citep{Horst18}, no lab experiments exist yet for hotter steam atmospheres. \\

In contrast to the active M-dwarf TRAPPIST-1, the host star in the L~98-59 system shows no evidence for stellar activity in the TESS data \citep{Kostov2019b} and is likely a relatively quiet M-dwarf with a low level of XUV activity. Such stars typically have an X-ray activity of ${\rm L_{x} \approx 10^{-4.5} L_{bol}}$ \citep{Shkolnik2014}. Using this activity level and scaling from the TRAPPIST-1 planet escape simulations of \citet{Bolmont2017}, we estimate that a steam atmosphere on L~98-59~b would lose $\sim 7$ Earth ocean equivalents per Gyr. This is orders of magnitude below the inferred water mass fraction of the planet --- assuming bulk composition similar to those previously inferred for the TRAPPIST-1 planets \citep{Grimm2018,Agol2021} --- illustrating that H$_2$O can still be present in the atmosphere of L~98-59~b, which has the highest anticipated transmission spectroscopy S/N of any planet smaller than 1.8 R$_\oplus$ of those discovered thus far. While L 98-59 b receives $\sim 23$ times Earth's incident flux, the detection of molecular absorption features on such a highly irradiated planet would be greatly significant as it would suggest that M-dwarf planets can readily retain their atmospheres. \\ 

Moreover, terrestrial planet formation simulations predict a wide variety of possible water abundances of up to many thousands of Earth oceans, and water mass fractions of $\sim$1-10\% --- particularly for systems like the planets of L~98-59 and TRAPPIST-1 that are close to mean motion resonances around an M dwarf, are suggestive of planet migration \citep[e.g.][]{Chambers2001,Hansen2012,Mordasini2012paperII, Tian2015,Unterborn2018}. The recent update in transit timing variations measured for the TRAPPIST-1 system \citep{Agol2021} has strongly revised the estimated water mass fractions for the inner planets b, c, and d, from $\gtrsim$5\% \citep{Grimm2018} down to $\lesssim$0.01\%, several times lower than Earth's ocean water mass fraction. However, this estimation is still consistent with significant water vapor in the planetary atmospheres.\\

\begin{figure*}[t]
    \centering
    \resizebox{12cm}{!}{\includegraphics{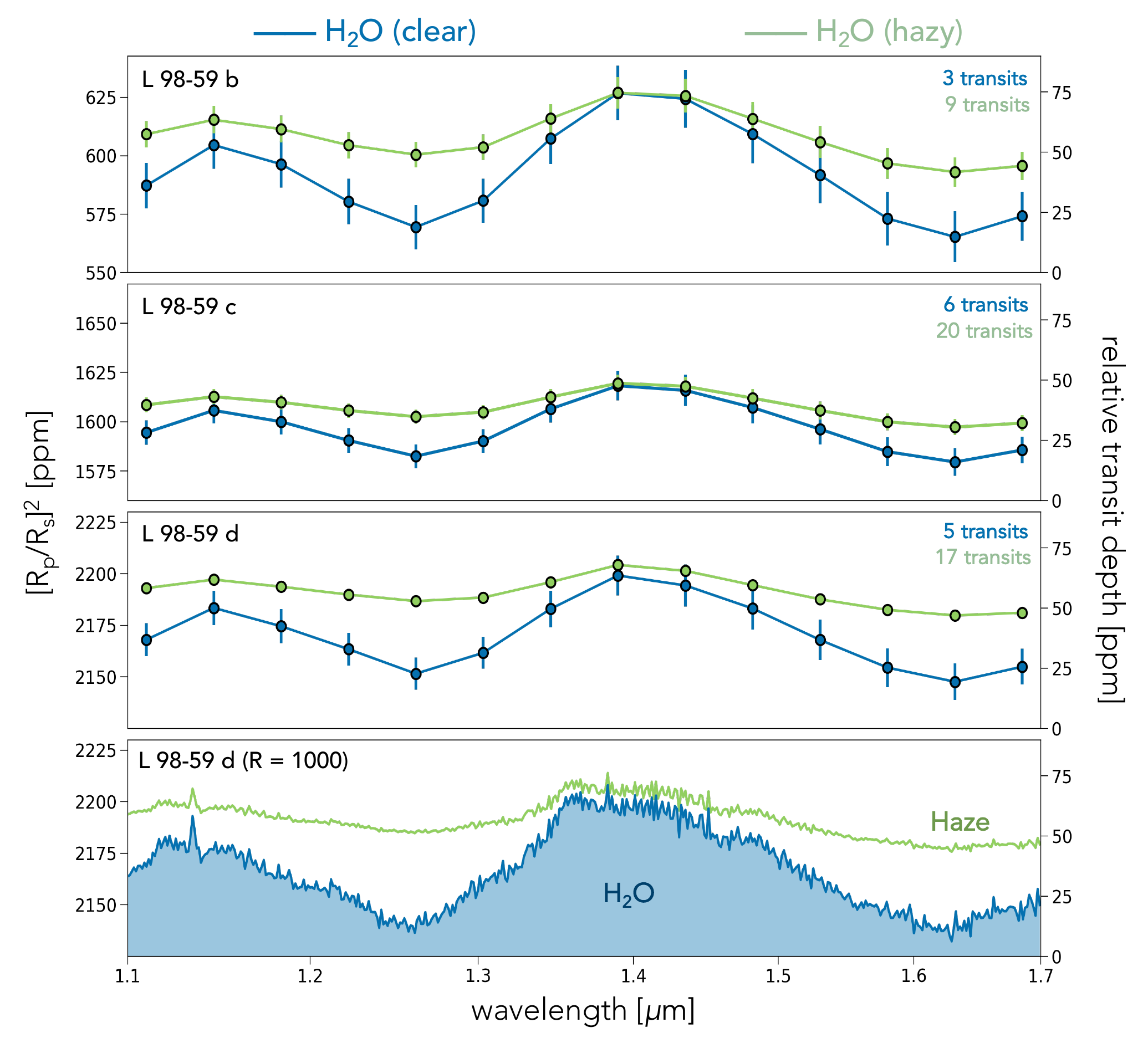}}\caption{Simulated HST/WFC3 transmission spectra of L 98-59 b, c, and d for a clear steam atmosphere (blue) and a hazy steam atmosphere (green). Error bars are shown for observations simulated at R = 30. The gaseous species corresponding to the molecular features in each hazy spectrum are shown in the bottom panel at a higher (R = 1000) resolution. Here, the green line represents the corresponding hazy simulated spectrum in panel C (green), while the blue indicates the absorption of H$_2$O. In this wavelength region, N$_2$ absorption impacts scale height rather than spectral features, so it is not shown. The number of transits simulated for each planet is set to obtain a 5$\sigma$ detection of H$_2$O at $1.38\ \mu$m. For the clear case (blue), this requires 3, 6, and 5 transits; while the hazy steam atmosphere requires 9, 20, and 17 transits for L 98-59 b, c, and d; respectively.}
    \label{fig:hst-water}
\end{figure*}

Figure \ref{fig:hst-water} shows model spectra of both a clear and hazy steam atmosphere with simulated HST/WFC3 observations for L 98-59 b, L 98-59 c, and L 98-59 d. The bottom panel accompanies these spectra to show the detailed molecular composition of each hazy steam atmosphere, using L 98-59 d as an example. The blue models show a clear sky spectrum with large H$_2$O absorption features and data uncertainties calculated for a 5$\sigma$ detection of the strongest spectral feature. The green models include Titan-based organic haze whose refractive indices are derived from \cite{Khare1984}, with data uncertainties calculated for a 5$\sigma$ detection of the same spectral feature. 

\ 

The clear steam atmosphere (blue) has a strong H$_2$O-absorption feature at 1.38 $\mu$m that can be detected at 5$\sigma$ in 3, 6, and 5 transits for L 98-59 b, c, and d; respectively. However, when considering the presence of a Titan-like haze (green), the number of transits required to detect the same 1.38 $\mu$m absorption feature at 5$\sigma$ more than doubles for each planet, requiring 9, 20, and 17 transits for L 98-59 b, c, and d; respectively. This decrease in sensitivity can be attributed to the continuum level in our modeled spectra which is set by Titan's high-altitude haze \citep{Robinson2014b}, whose placement will have more dramatic effects at UV/VIS/NIR wavelengths, where the haze particles are strongly absorbing \citep{Tomasko2008}. The increase in haze production as we move out of the system indicates a link between aerosol formation and equilibrium temperature predicted by theoretical models for warm-Neptune and super-Earth planets \citep{Fortney2013,morley2015}. \cite{Crossfield2017} hypothesize that the size of the water absorption feature of an exoplanet in the WFC3 bandpass may be related to the equilibrium temperature of the planet. When considering warm-Neptunes, hotter planets appear to have larger observable features, whereas cooler planets are able to more efficiently form high-altitude aerosol layers, muting these absorption features. The resulting flattened spectra that can be attributed to high-altitude aerosol layers have been observed with WFC3 for large-mass low-density planets \citep{LibbyRoberts2020}, but additional observations will be required to confirm or refute this trend for both mini-Neptune and terrestrial planets.

\begin{table}[H]
\centering
\caption{The modeled observability of HST within this study}
\begin{tabular}{c   c    c  c}
\hline
\hline
Planet & L 98-59 b & L 98-59 c & L 98-59 d\\
\hline                                              
Instrument & \multicolumn{3}{c}{WFC3 (R = 30)}\\
\hline
\hline
Atmosphere & \multicolumn{3}{c}{H$_2$/He-dominated}\\
\hline 
Spectral feature & \multicolumn{3}{c}{H$_2$O $1.38\ \mu$m} \\
(S/N)-1 & --- & 10 & 13 \\
N transits ($5\sigma$) & --- & 1 & 1 \\
\hline 
Spectral feature & \multicolumn{3}{c}{CH$_4$ $1.68\ \mu$m} \\ 
(S/N)-1 & --- & 11 & 13 \\
N transits ($5\sigma$) & --- & 1 & 1 \\
\hline
\hline
Atmosphere & \multicolumn{3}{c}{Steam} \\ 
\hline
Spectral feature & \multicolumn{3}{c}{H$_2$O $1.38\ \mu$m; Clear} \\
(S/N)-1  & 3.0 & 2.1 & 2.0  \\
N transits  ($5\sigma$) & 3 & 6 & 5   \\
\hline
Spectral feature & \multicolumn{3}{c}{H$_2$O $1.38\ \mu$m; Hazy} \\
(S/N)-1 & 1.7 & 1.1 & 1.2 \\ 
N transits (5$\sigma$) & 9 & 20 & 17 \\
\hline 
\label{tab:hstresults}
\end{tabular}
\end{table}


\section{Prospects and Results for Future Characterization with JWST} 
\label{sec:jwst}
L 98-59 will be one of the first planetary systems characterized by JWST. JWST's Guaranteed Time Observations (GTO) include transits for L 98-59 c (NIRISS/SOSS, 1 transit, program number 1201, PI Lafreniere) and L 98-59 d (NIRSpec/G395H, 2 transits, program number 1224, PI Birkmann; NIRISS/SOSS, 1 transit, program 1201, PI Lafreniere). Therefore, simulated transit observations of the L 98-59 planets using these JWST instruments are timely. If HST observations indicate that the planets in the L 98-59 system are absent of low mean-molecular weight gases, this motivates the follow-up search for secondary atmospheres dominated by high mean-molecular weight species such as CO$_2$ or O$_2$ to be the primary constituent of some or all of the planetary atmospheres in this system. \\ 

In the following subsections, we present model transmission spectra of each planet using NIRISS SOSS, NIRSpec/G395H, and MIRI LRS. Details on S/N and transit values for individual spectral features are found in Table \ref{tab:jwstresults}. 
\subsection{{\texorpdfstring{Detecting H\textsubscript{2}}OO with JWST}}
Although HST may be able to detect the presence of a clear, water-dominated atmosphere on each planet in six transits or less, the addition of Titan-like organic aerosols significantly raises the continuum level between 1.1 - 1.7 $\mu$m, decreasing the sensitivity of the H$_2$O absorption feature at 1.38 $\mu$m and tripling the number of transits required for detection (Figure \ref{fig:hst-water}). Knowing that the haze particles most strongly absorb at UV/VIS/NIR wavelengths \citep{Tomasko2008}, we revisit this atmospheric scenario with the increased wavelength range of JWST's NIRISSS instrument. \\

\begin{figure*}[t]
    \centering
    \resizebox{18.5cm}{!}{\includegraphics{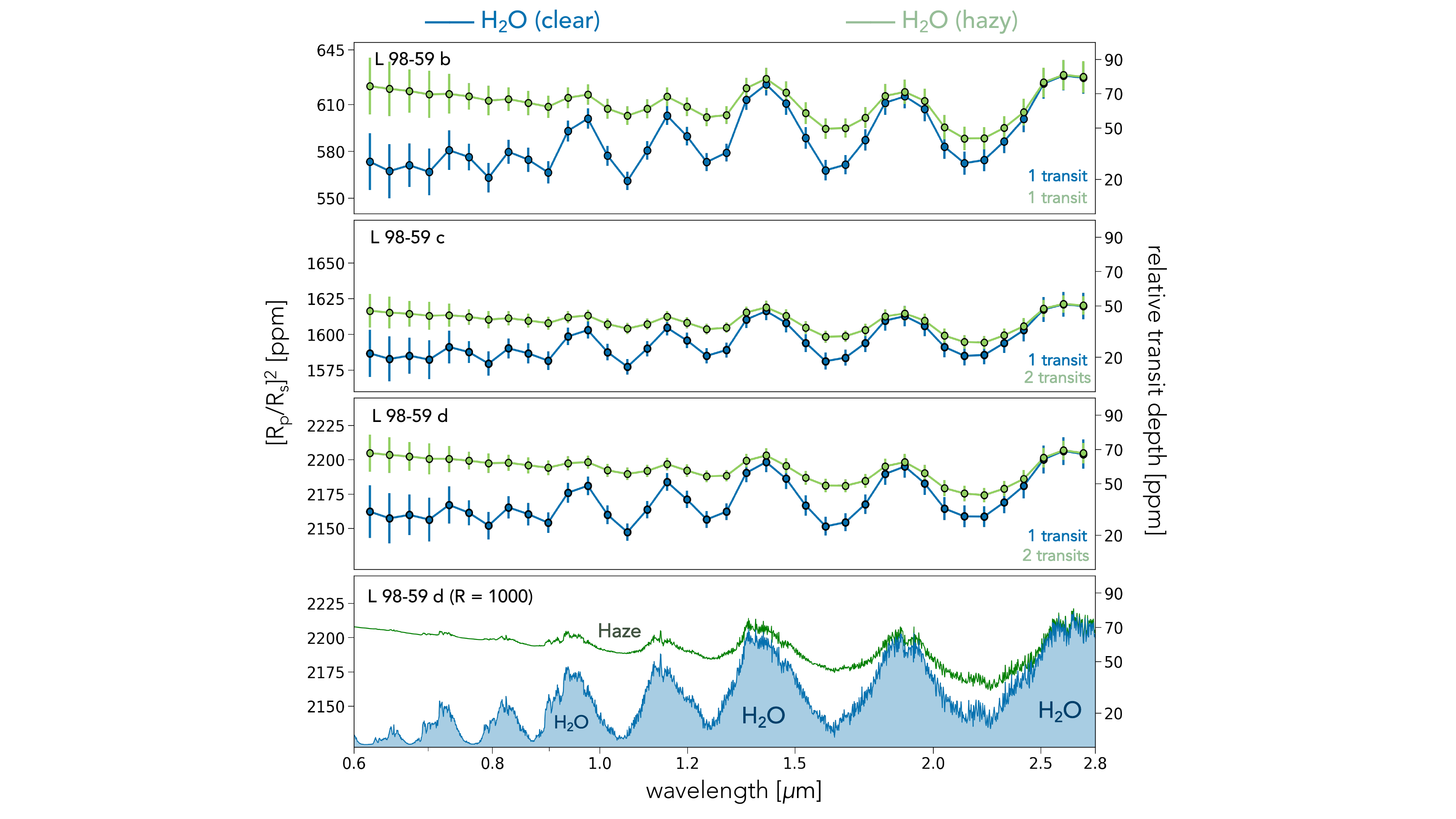}}
    \caption{Simulated JWST/NIRISS SOSS transmission spectra of L 98-59 b, c, and d for a clear steam atmosphere (blue) and a hazy steam atmosphere (green). Error bars are shown for observations simulated at R = 30. The gaseous species corresponding to the molecular features in each hazy spectrum are shown in the bottom panel at a higher (R = 1000) resolution. Here, the green line represents the corresponding hazy simulated spectrum in panel C (green), while the blue indicates the absorption of H$_2$O. In this wavelength region, N$_2$ absorption impacts scale height rather than spectral features so it is not shown. The number of transits simulated for each planet is set to obtain a 5$\sigma$ detection of H$_2$O at $1.38\ \mu$m. For the clear case, this requires 1 transit for each planet, while the hazy case requires 1, 2, and 2 transits for L 98-59 b, c, and d; respectively.}
    \label{fig:jwst-water}
\end{figure*}

Figure \ref{fig:jwst-water} shows model spectra of both a clear and hazy steam atmosphere (the same as shown in Figure \ref{fig:hst-water}) with simulated JWST/NIRISS SOSS observations for L 98-59 b,  L 98-59 c, and L 98-59 d. The bottom panel accompanies these spectra to show the detailed molecular composition of each hazy steam atmosphere, using L 98-59 d as an example. The blue models show a clear sky spectrum with large H$_2$O absorption features and data uncertainties calculated for a 5$\sigma$ detection of the strongest spectral feature. The green models include Titan-based organic haze whose refractive indices are derived from \cite{Khare1984}, with data uncertainties calculated for a 5$\sigma$ detection of the same spectral feature. 

For a clear steam atmosphere (blue), NIRISS SOSS may be able to rule out a featureless spectrum at 5$\sigma$ in $\leq$ 1 transit for all three planets. Although NIRISS's wavelength range (0.6 - 2.8 $\mu$m) is much more extensive than that of HST/WFC3/G141 (1.1 - 1.7 $\mu$m), H$_2$O absorption at 1.38 $\mu$m is still calculated to be the most detectable feature in the spectrum, providing a S/N $\geq$ 5 in exchange for nominal observational time. When considering the presence of a Titan-like haze (green), the number of transits required to detect the same 1.38 $\mu$m absorption feature with the same confidence level increases to 1, 2, and 2 transits for L 98-59 b, c, and d; respectively. This drop in sensitivity is much less dramatic than that seen in HST/WFC3 (Figure \ref{fig:hst-water}), and can be attributed to the haze opacity effects in transit that become negligible in the midinfrared (MIR) \citep{Robinson2014b}. As the haze continuum slope is strongly wavelength dependent, refraction and gas absorption become responsible for distinguishing spectral features beginning around 2.5 $\mu$m, where the magnitude of the transit height variations caused by the haze continuum are no longer equally as large as the observed gaseous features. Having access to redder NIR wavelengths also allows us to probe for other H$_2$O absorption features (such as that at 2.6 $\mu$m) that are not diminished by the presence of haze. This multiband detection would increase the certainty level of our observations, which is a critical step in constraining the chemical composition of planetary atmospheres such as these.

\subsection{{\texorpdfstring{Detecting CO\textsubscript{2}}w with JWST}}
One important question to answer is whether or not CO$_2$ is a dominant gas in the atmospheres of any of these planets. Amongst the three rocky planets in our solar system with substantial atmospheres, CO$_2$ dominates two of their atmospheres (Venus and Mars), and is thought to have been a prominent gas in Earth's early atmosphere during the Hadean epoch \citep{Zahnle2011}. Emerging evidence also suggests that CO$_2$ was substantially more abundant during the Archean \citep{CatlingZahnle2020}. This predominance in the solar system suggests CO$_2$ atmospheres could be common in other planetary systems as well. Nevertheless, even at the relatively low abundances of today, CO$_2$ is a critical greenhouse gas that maintains planetary temperatures and generates substantial spectral features.  \\ 

CO$_2$ has a strong opacity near 4.3 $\mu$m, allowing for the possibility of achieving a $5\sigma$ detection with JWST, as shown by \citet{Morley2017,Lincowski2018,KrissansenTotton2018,Fauchez2019,Wunderlich2019,Lustig-Yaeger2019,Pidhorodetska2020} for a variety of modeled atmospheric scenarios when considering the TRAPPIST-1 habitable zone planets. For a post-runaway planet like Venus in our solar system, the detection of CO$_{2}$ at $4.3 \ \mu$m will provide a key indicator. CO$_{2}$ is well-mixed in the atmosphere and therefore much less sensitive to thick clouds more likely to be found in the troposphere \citep{robinson2014}, and its high mean-molecular weight aids in the survival of potential atmospheric escape. In addition, a Venus-like atmosphere is less sensitive to changes in temperature due to the low condensation temperature of CO$_2$ \citep{Morley2017}. \\ 

\begin{figure*}[t]
    \centering
    \resizebox{13cm}{!}{\includegraphics{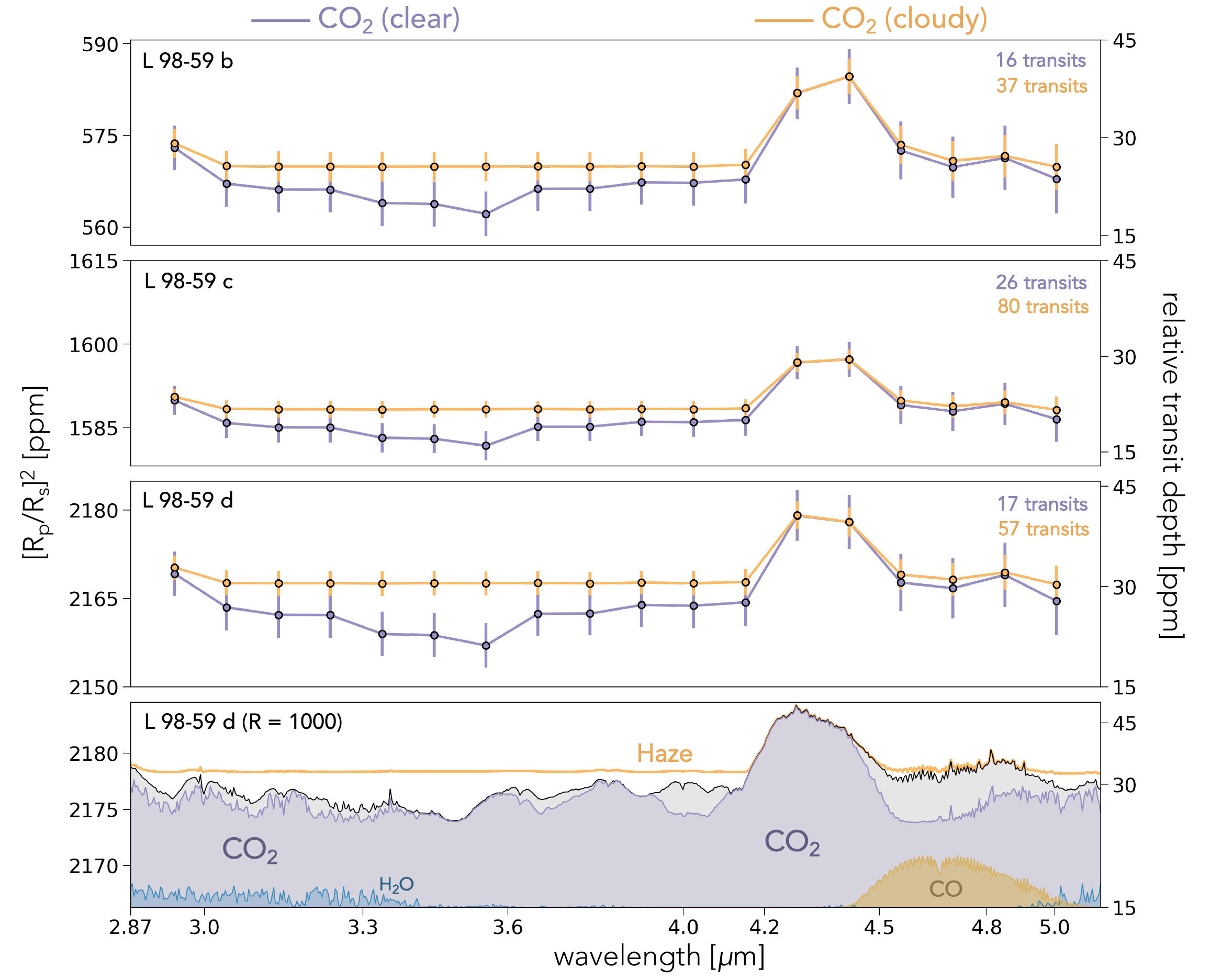}}
    \caption{Simulated JWST/NIRSpec G395H transmission spectra of L 98-59 b, c, and d for a clear CO$_2$-dominated atmosphere (purple) and a CO$_2$-dominated atmosphere that includes Venus-like H$_2$SO$_4$ clouds (orange). Error bars are shown for observations simulated at R = 30. The bottom panel corresponds to the molecular composition of the cloudy CO$_2$ atmosphere for L 98-59 d shown at a higher (R = 1000) resolution. Here, the black line represents the clear simulated spectrum in panel C, the orange line represents the cloudy simulated spectrum in panel C, and the grey indicates the sum of all gases present. The absorption of CO$_2$ is shown in purple, H$_2$O in blue, and CO in yellow. The number of transits simulated for each planet is set to obtain a 5$\sigma$ detection of CO$_2$ at $4.3\ \mu$m. For the clear case, this requires 16, 26, and 17 transits; while the cloudy case requires 37, 80, and 57 transits for L 98-59 b, c, and d; respectively.}
    \label{fig:jwst-co2-clear}
\end{figure*}

Figure \ref{fig:jwst-co2-clear} shows model spectra of both clear and cloudy Venus-like atmospheres with simulated JWST/NIRSpec G395H observations for L 98-59 b, L 98-59 c, and L 98-59 d. The bottom panel accompanies these spectra to show the detailed molecular composition of each cloudy atmosphere, using L 98-59 d as an example. The purple models show a clear sky spectrum with large CO$_2$ absorption features and data uncertainties calculated for a 5$\sigma$ detection of the strongest spectral feature. The orange models include H$_2$SO$_4$ clouds whose refractive indices are derived from \cite{palmer1975}, with data uncertainties calculated for a 5$\sigma$ detection of the same spectral feature. 

\ 

The clear Venus-like atmosphere (purple) shows multiple spectral features (such as at 2.8 $\mu$m) that can be attributed to CO$_2$ absorption, but the detectability of this atmosphere is dominated by the strength of the 4.3 $\mu$m CO$_2$ absorption feature. In this cloud-free scenario, a featureless spectrum could be ruled out at 5$\sigma$ in 16, 26, and 17 transits for L 98-59 b, c, and d; respectively. This requires considerably fewer transits than the cloudy Venus-like atmosphere (orange), where the H$_2$SO$_4$ clouds are located high enough in the atmosphere that a detection at the same confidence level would involve 37, 80, and 57 transits for L 98-59 b, c, and d; respectively. Contrary to the Titan-like organic haze (Figure \ref{fig:jwst-water}), the continuum level set by the H$_2$SO$_4$ clouds is not wavelength dependent in the NIRSpec G395H range, causing considerable flattening across the entire spectrum. We note this same effect in the MIRI LRS wavelength range. 

\subsection{{\texorpdfstring{Detecting O\textsubscript{2}}w with JWST}}
Desiccated O$_2$ atmospheres have been suggested as one possible outcome of a prolonged runaway greenhouse phase combined with extensive water loss, where all of the oceans in an atmosphere evaporate. Exposed to UV radiation,  H$_2$O breaks into O and H, where the H will then escape. The remaining composition is then dominated by photochemically-produced O$_2$ at least temporarily. Photochemically produced O$_2$ may then be consumed by a magma ocean. The extent to which O$_2$ will be left behind in the atmosphere is dependent on a multitude of factors such as the longevity of the magma ocean, the extent of water loss (and therefore O$_2$ buildup), ameliorating effects of O$_2$ loss via hydrodynamic escape, and subsequent reactions with reducing volcanic gases over time. Such an atmospheric composition  may be common for planets around M-type stars \citep{Tian2015,LugerBarnes2015,Schwieterman2016,Meadows2017,Meadows2018b,Lustig-Yaeger2019}. In addition, the molecular weight of O$_2$ makes it more resistant to hydrodynamic escape than less massive homonuclear molecules such as H$_2$. \\ 

For the TRAPPIST-1 system, \cite{Lincowski2018} have shown that for an assumed original water content of 20 Earth oceans (by mass), TRAPPIST-1b, c and d may have lost all their water while TRAPPIST-1e, f and g may have lost between 3 to 6 Earth oceans. As a result, the atmospheres of these planets may have accumulated the amount of O$_2$ that would be equivalent to surface pressures between 22 and 5,000 bars, if all of the O$_2$ were to remain in the atmosphere. This mechanism can be limited by oxygen absorption into a magma ocean if temperature conditions are high enough \citep{Wordsworth_2018}. However, if the mantle of these planets becomes oxidized, some amount of O$_2$ could still persist in the atmosphere over geologic timescales. Therefore, it is possible that large amounts of O$_2$ can still be present in the atmospheres of the TRAPPIST-1 planets (including the innermost), depending on the degree of oxidation of the mantle. This creates a unique opportunity to explore similar cases within the L 98-59 system, providing an avenue for comparative planetology in addition to atmospheric characterization.  \\ 

Molecular O$_2$ absorbs through vibrational and rotational transitions at several wavelengths in the visible (VIS) at 0.63, 0.69, and 0.76 $\mu$m, and in the near-infrared (NIR) at 1.27 $\mu$m. In addition, O$_2$-O$_2$ CIA produces broad spectral features distinct from individual O$_2$ absorption features at UV, Visible, and near-IR wavelengths. Two O$_2$-O$_2$ CIA features are present in the HST/WFC3 wavelength range at 1.06 and 1.27 $\mu$m. Although the sensitivity of HST/WFC3 is not strong enough to detect these features on terrestrial planets, JWST's NIRISS could be sensitive enough to probe for these gases. \cite{Schwieterman2016} proposed that these two CIA features could be used in transmission spectroscopy to identify abiotically produced desiccated and dense O$_2$ atmospheres, building on early work by \cite{Misra2014} that proposed the use of near-IR O$_2$-O$_2$ CIA features as a method for measuring minimum atmospheric pressure. \cite{Lustig-Yaeger2019} have studied this possibility for the TRAPPIST-1 system assuming between 10 to 100 bars of O$_2$ in the atmospheres of the planets and have shown that JWST can detect these O$_2$-O$_2$ CIA features at 5$\sigma$ in just a few transits. In addition, \cite{Fauchez2019b} have demonstrated that O$_2$ in low-pressure atmospheres can be detected with MIRI using O$_2$-X CIA (X can be any gas) at $6.4\ \mu$m, assuming no confounding impacts of large H$_2$O abundances. \\

When modeling a post-runaway desiccated planet that is rich in abiotic O$_2$, we expect to see O$_3$ formation from the photochemical processing of this O$_2$. The formation and abundance of ozone in a planetary atmosphere will be strongly coupled to its temperature structure. To first order, the formation and destruction in Earth’s atmosphere is controlled by the Chapman reaction scheme \citep{Chapman1930}. We write this series of reactions below:
\begin{equation}\label{eq:chapman1}
    \mathrm{O}_{2}+\mathrm{hc} / \lambda(\lambda<242 \mathrm{nm}) \rightarrow \mathrm{O}+\mathrm{O}
\end{equation}
\begin{equation}\label{eq:chapman2}
    \mathrm{O}+\mathrm{O}_{2}+\mathrm{M} \rightarrow \mathrm{O}_{3}+\mathrm{M}
\end{equation}
\begin{equation}\label{eq:chapman3}
    \mathrm{O}_{3}+\mathrm{O} \rightarrow \mathrm{O}_{2}+\mathrm{O}_{2}
\end{equation} \ 

Only one of these reactions (Equation (\ref{eq:chapman2})) produces ozone. The rate constant of Equation (\ref{eq:chapman2}) in the low-pressure limit, which is practically the case throughout the atmosphere, is given as K$_{(5)}$ = 6.1$\times$10$^{-34}$ (298/T)$^{2.4}$ cm$^{6}$ molecules$^{-2}$ s$^{-1}$ \citep{Burkholder2019}. This rate constant is geometrically inversely proportional to the temperature of the atmosphere in which the reaction is occurring. This is because the reaction is really a set of three tightly coupled reactions: 

\begin{equation}
    \mathrm{O}_{2}+\mathrm{O} \rightarrow \mathrm{O}_{3}* \tag{5.1} \label{eq:5.1}
\end{equation}
\begin{equation}
    \mathrm{O}_{3}* \rightarrow \mathrm{O}_{2}+\mathrm{O} \tag{5.2}
    \label{eq:5.2}
\end{equation}
\begin{equation}
    \mathrm{O}_{3}* +\mathrm{M} \rightarrow \mathrm{O}_{3}+ \mathrm{M} \tag{5.3} \label{eq:5.3} 
\end{equation}

Where O$_3$* is not a distinct species but rather an excited state of O$_2$ and O in a collisional but unbound state. Without an inert collision to carry away excess energy, the O$_3$ decays into O$_2$ (O$_3$* $\rightarrow$ O$_2$ + O). The lifetime of this state is strongly inversely related to temperature as higher temperatures mean O$_2$ and O are in proximity for a shorter amount of time and there is less opportunity for M to take away the excess energy. Therefore, we would strongly anticipate that hot planetary atmospheres (T$>$300~K) would suppress ozone formation relative to cooler atmospheres. Note that this simplified description does not consider additional catalytic pathways to ozone destruction (e.g. Cl, NO$\subtxt{x}$, Br, etc.) or atmospheric transport. However, consideration of these pathways would lead to even lower estimates for O$_3$ prevalence. \\ 

\begin{figure*}[t]
    \centering
    \resizebox{16cm}{!}{\includegraphics{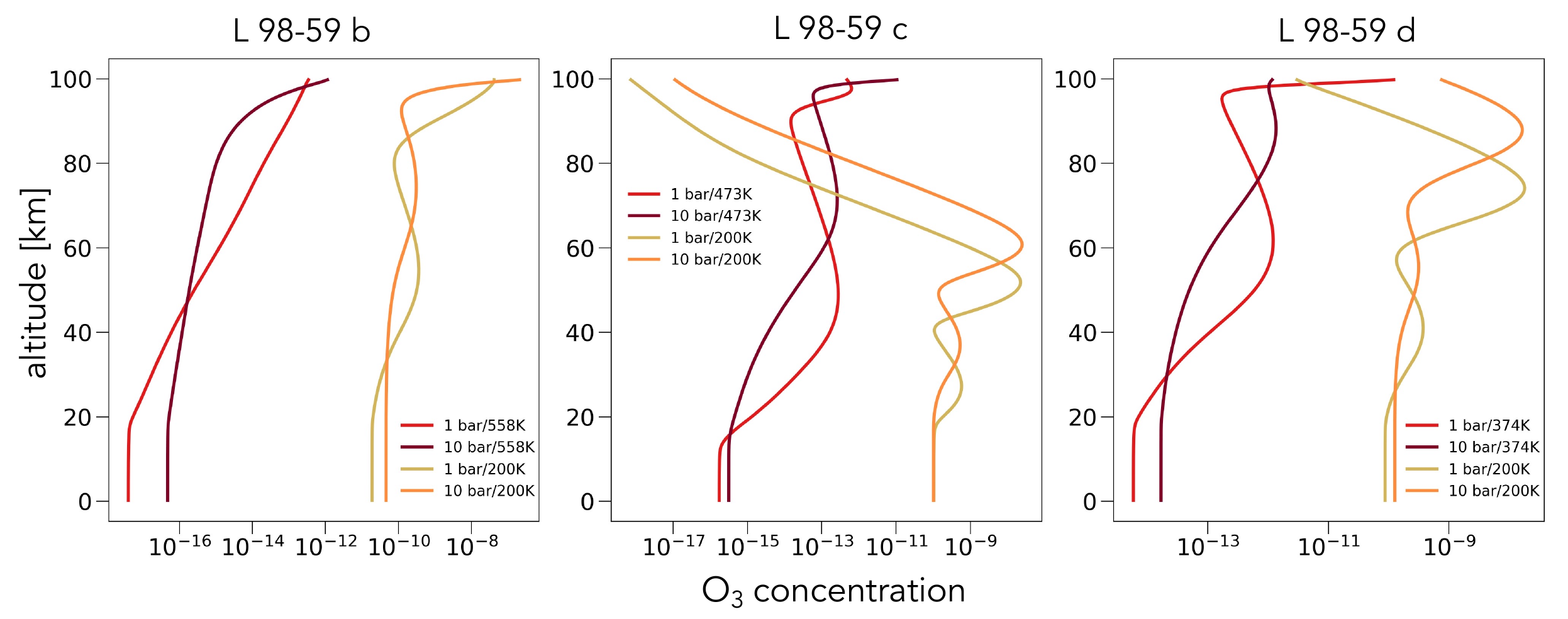}}
    \caption{O$_3$ concentrations for each modeled O$_2$/O$_3$ atmosphere as a function of altitude [km]. These outputs are obtained from the Atmos model.}
    \label{fig:o3-concentration}
\end{figure*}

Each of the planets in the L 98-59 system have high equilibrium temperatures, though the temperature structures of their atmospheres are unknown and will depend on the presence or absence of stratospheric absorbers and line cooling by molecular species other than O$_2$. We bound this problem by using Atmos \citep{Arney2016} to model ozone production in 1 and 10 bar O$_2$-dominated atmosphere scenarios both for an isothermal temperature of 200K and an isothermal temperature equal to the equilibrium temperature of the planet (b = 558K, c = 473K, d = 374K). This approach is similar to that previously employed by \cite{Fauchez2019b} when estimating the possible range of scale heights relevant to transmission spectroscopy in O$_2$-dominated atmospheres, except that we also model O$_3$ photochemistry. The O$_3$ outputs from the Atmos model are detailed in Figure \ref{fig:o3-concentration}. 

\ 

\begin{figure*}[t]
    \centering
    \resizebox{17cm}{!}{\includegraphics{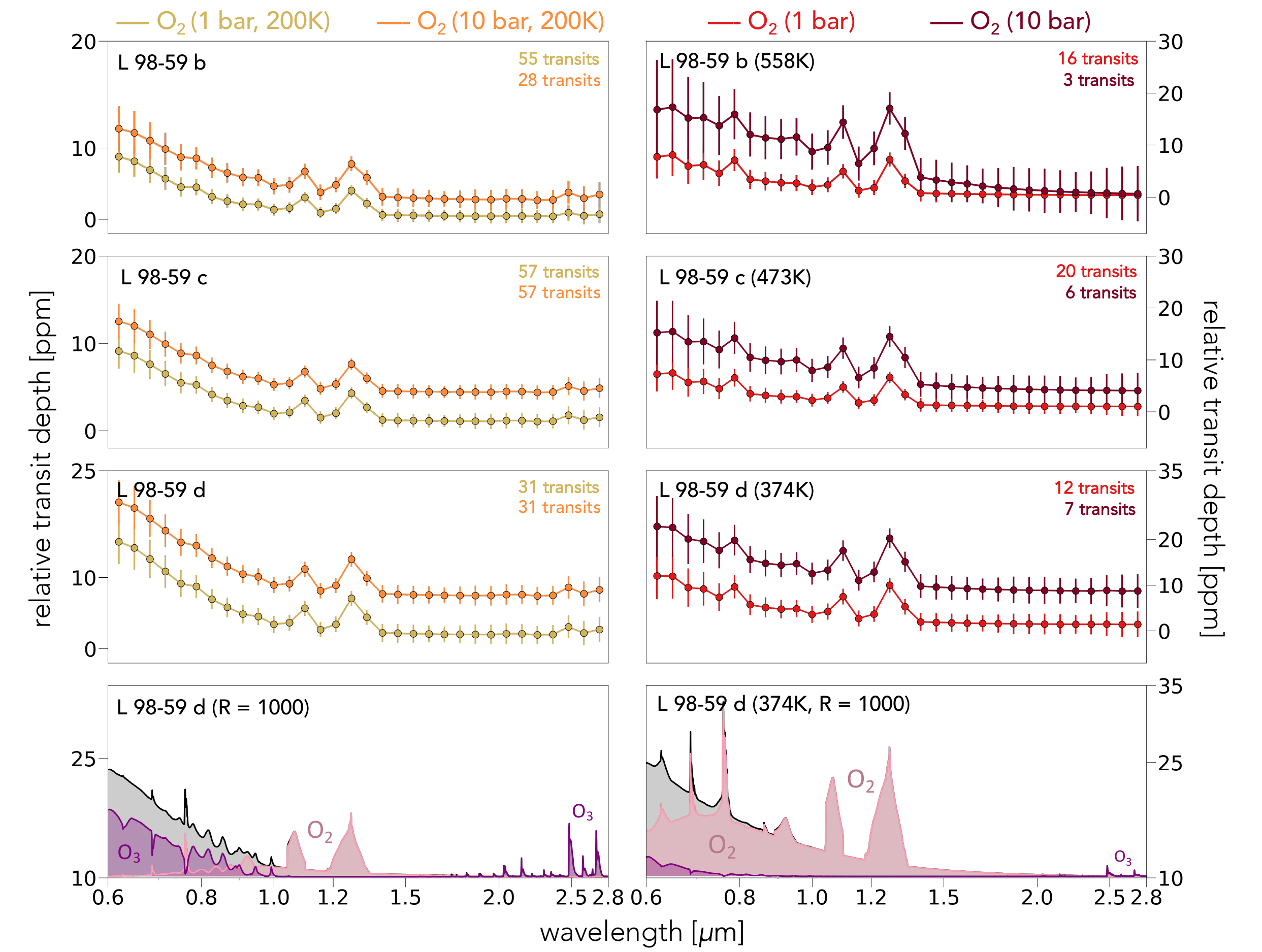}}
    \caption{Simulated JWST/NIRISS SOSS transmission spectra of L 98-59 b, c, and d with varying equilibrium temperatures and surface pressures for four possible O$_2$/O$_3$ atmospheric states. The left column shows 200K for each planet at 1 bar (yellow) and 10 bar (orange), while the right column shows the equilibrium temperature for each planet at 1 bar (red) and 10 bar (burgundy). Error bars are shown for observations simulated at R = 20. The bottom panel in each column corresponds to the molecular composition of each 10 bar atmosphere for L 98-59 d, which consists entirely of O$_2$ (pink) and O$_3$ (purple). The number of transits simulated for each planet is set to obtain a 5$\sigma$ detection of O$_2$ at 1.27 $\mu$m, with values listed in Table \ref{tab:jwstresults}.}
    \label{fig:jwst-o2-nirspec}
\end{figure*}

Figure \ref{fig:jwst-o2-nirspec} shows model transmission spectra representing a variation of four potential O$_2$ atmospheres for L 98-59 b, c, and d with NIRISS SOSS. The left column shows spectra for 200K at 1 bar (yellow) and 10 bar (orange), while the right column shows spectra for each planet's equilibrium temperature at 1 bar (red) and 10 bar (burgundy). The bottom panel accompanies these spectra to show the detailed molecular composition of each 10 bar atmosphere for L 98-59 d.

\ 

We find that if the L 98-59 planets possess stratospheric temperatures that are significantly warmer than Earth's ($\sim200-270$K), ozone formation will be suppressed even if their atmospheres are completely dominated by (abiotic) oxygen. At their native equilibrium temperatures (Figure \ref{fig:jwst-o2-nirspec} right column), the 1.27 $\mu$m O$_2$ absorption band could be detected at 5$\sigma$ with 16, 20, and 12 transits (1 bar) and 3, 6, and 7 transits (10 bar) for L 98-59 b, c, and d; respectively. However, at a cooler equilibrium temperature of 200K (Figure \ref{fig:jwst-o2-nirspec} left column), ozone formation increases. In this scenario, the direct detection of O$_3$ absorption could be another key indicator of a desiccated atmosphere, but we find that the decrease in scale height with temperature renders the various O$_3$ features from 2-3 $\mu$m to be undetectable, requiring hundreds of transits for a 5$\sigma$ detection. The strength of O$_2$ absorption at 1.27 $\mu$m is also significantly diminished, requiring 55, 57, and 31 transits (1 bar) and 28, 57, and 31 transits (10 bar) for L 98-59 b, c, and d; respectively. In summary, at low temperatures, ozone formation is more efficient (see bottom left panel of Figure \ref{fig:jwst-o2-nirspec}) but small scale heights impede ozone detectability. In contrast, at high temperatures the formation of ozone is suppressed but this is more than compensated for by increased scale height.

\ 

\begin{figure*}[t]
    \centering
    \resizebox{15cm}{!}{\includegraphics{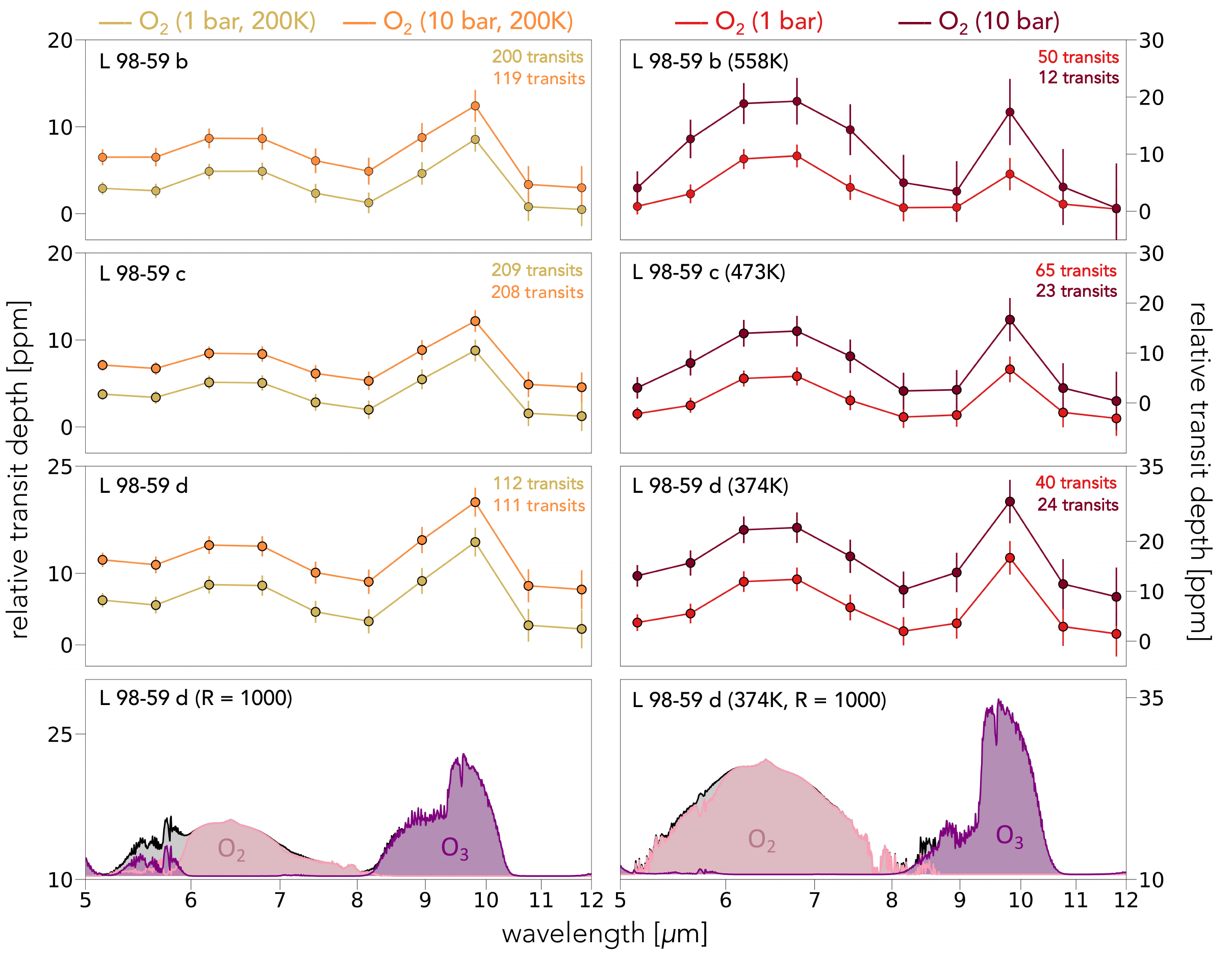}}
    \caption{Simulated JWST/MIRI LRS transmission spectra of L 98-59 b, c, and d with varying equilibrium temperatures and surface pressures for four possible O$_2$/O$_3$ atmospheric states. The left column shows 200K for each planet at 1 bar (yellow) and 10 bar (orange), while the right column shows the equilibrium temperature for each planet at 1 bar (red) and 10 bar (burgundy). Error bars are shown for observations simulated at R = 10. The bottom panel in each column corresponds to the molecular composition of each 10 bar atmosphere for L 98-59 d, which consists entirely of O$_2$ (pink) and O$_3$ (purple). The number of transits simulated for each planet is set to obtain a 5$\sigma$ detection of O$_2$ at 6.4 $\mu$m, with values listed in Table \ref{tab:jwstresults}.}
    \label{fig:jwst-o2-miri}
\end{figure*}

Beyond the wavelength range of NIRSpec, transit observations with MIRI LRS would allow us to search for O$_2$/O$_3$ absorption features between 5-12 $\mu$m. Figure \ref{fig:jwst-o2-miri} shows model spectra of the same atmospheres in Figure \ref{fig:jwst-o2-nirspec} for L 98-59 b, c, and d with MIRI LRS. For all planets, the broad O$_2$-O$_2$ CIA band centered at 6.4 $\mu$m becomes increasingly detectable as atmospheric temperatures and pressures are raised. At their native equilibrium temperatures (Figure \ref{fig:jwst-o2-miri} right column), O$_2$ could be detected at 5$\sigma$ with 50, 65, and 40 transits (1 bar) and 12, 23, and 24 transits (10 bar) for L 98-59 b, c, and d; respectively. However, at a cooler equilibrium temperature of 200K (Figure \ref{fig:jwst-o2-miri} left column), targeting this feature for a 5$\sigma$ detection would require upwards of 100 transits for each planet, at both 1 and 10 bar surface pressures. A direct detection of O$_3$ could help constrain atmospheric conditions if O$_2$ is rendered undetectable, but we find that the O$_3$ absorption feature at 9.7 $\mu$m is heavily impacted by noise in the infrared, and would require $\geq$ 200 transits for each of the L 98-59 planets at T$\subtxt{eq}$ = 200K. At their native equilibrium temperatures, less ozone is formed, but the increase in scale height may allow for O$_3$ detection in atmospheres with surface pressures of 10 bar or higher, requiring 36, 41, and 32 transits for L 98-59 b, c, and d; respectively. We find that transit observations with NIRISS SOSS are more optimal than MIRI LRS as they require significantly fewer transits to detect O$_2$ absorption features. 

\begin{table*}[t]
\centering
\caption{JWST instruments and their observability in this study}
\begin{tabular}{c c c c} 
\hline
\hline
 Planet & L 98-59 b & L 98-59 c & L 98-59 d\\
\hline 
Instrument & \multicolumn{3}{c}{NIRSpec G395H}\\
\hline 
\hline 
Atmosphere & \multicolumn{3}{c}{Venus-like}
(R = 30) \\ 
\hline 
Spectral feature & \multicolumn{3}{c}{CO$_2$ $4.3\ \mu$m; Clear} \\
(S/N)-1 & 1.3 & 0.98 & 1.2   \\
N transits  ($5\sigma$) & 16 & 26 & 17 \\
\hline
Spectral feature & \multicolumn{3}{c}{CO$_2$ $4.3\ \mu$m; Cloudy} \\
(S/N)-1 & 0.83 & 0.56 & 0.66   \\
N transits  ($5\sigma$) & 37 & 80 & 57 \\
\hline
\hline
Instrument & \multicolumn{3}{c}{NIRISS SOSS}\\
\hline
\hline
Atmosphere & \multicolumn{3}{c}{Steam}
(R = 23) \\ 
\hline 
Spectral feature & \multicolumn{3}{c}{H$_2$O $1.38\ \mu$m; Clear} \\
(S/N)-1 & 8.7 & 6.1 & 6.8  \\
N transits  ($5\sigma$) & 1 & 1 & 1   \\
\hline
Spectral feature & \multicolumn{3}{c}{H$_2$O $1.38\ \mu$m; Hazy} \\
(S/N)-1 & 5.4 & 3.8 & 3.9  \\
N transits  ($5\sigma$) & 1 & 2 & 2\\
\hline 
\hline
Atmosphere & \multicolumn{3}{c}{O$_2$/O$_3$-desiccated}
(R = 20) \\ 
\hline 
Spectral feature & \multicolumn{3}{c}{O$_2$ CIA $1.27\ \mu$m, 1 bar, 200K} \\
(S/N)-1 & 0.68 & 0.67 & 0.9  \\
N transits  ($5\sigma$) & 55 & 57 & 31  \\
\hline
Spectral feature & \multicolumn{3}{c}{O$_2$ CIA $1.27\ \mu$m, 10 bar, 200K} \\
(S/N)-1 & 0.95 & 0.67 & 0.9  \\
N transits  ($5\sigma$) & 28 & 57 & 31  \\
\hline
Spectral feature & \multicolumn{3}{c}{O$_2$ CIA $1.27\ \mu$m, 1 bar, Eq. Temp} \\
(S/N)-1 & 1.3 & 1.1 & 1.5  \\
N transits  ($5\sigma$) & 16 & 20 & 12  \\
\hline
Spectral feature & \multicolumn{3}{c}{O$_2$ CIA $1.27\ \mu$m, 10 bar, Eq. Temp} \\
(S/N)-1 & 3.0 & 2.1 & 2.0  \\
N transits  ($5\sigma$) & 3 & 6 & 7  \\
\hline 
\hline 
Instrument & \multicolumn{3}{c}{MIRI}\\
\hline
\hline
Atmosphere & \multicolumn{3}{c}{O$_2$/O$_3$-desiccated}
(R = 10) \\ 
\hline 
Spectral feature & \multicolumn{3}{c}{O$_2$ CIA $6.4\ \mu$m, 1 bar, Eq. Temp} \\
(S/N)-1 & 0.71 & 0.62 & 0.8  \\
N transits  ($5\sigma$) & 50 & 65 & 40 \\
\hline
Spectral feature & \multicolumn{3}{c}{O$_2$ CIA $6.4\ \mu$m, 10 bar, Eq. Temp} \\
(S/N)-1 & 1.5 & 1.1 & 1.0  \\
N transits  ($5\sigma$) & 12 & 23 & 24 \\
\hline
Spectral feature & \multicolumn{3}{c}{O$_3$ $9.7\ \mu$m, 10 bar, Eq. Temp} \\
(S/N)-1 & 0.84 & 0.78 & 0.9 \\
N transits ($5\sigma$) & 36 & 41 & 32  \\
\hline
\label{tab:jwstresults}
\end{tabular}
\end{table*}

\section{Discussion} \label{sec:discussion}
In order to prepare for the future observations of the L 98-59 system that will be carried out using HST and JWST, we performed a series of theoretical models to better understand the atmospheres of the planets in the system and to predict the outcome of the planned observations. We limited the analysis to focus on transmission spectroscopy with HST using WFC3 and JWST using NIRSpec G395H, NIRISS SOSS, and MIRI LRS. Our sample consisted of all three planets in the system, with the two innermost being almost certainly rocky, while the outermost is likely a mini-Neptune. These simulations varied the equilibrium temperature, cloud coverage, and overall atmospheric composition. However, they did not represent the full spectrum of possible outcomes. The results presented herein result from some optimistic assumptions regarding planetary parameters, and therefore represent lower limits on the amount of observing time needed to detect and characterize atmospheres in the L 98-59 system. As a result, observing plans that include fewer transits than reported here may require additional observations to make robust detections on the nature of these worlds. We now address the influence of these assumptions along with further considerations that would broaden our understanding of our conclusions. 

\subsection{Atmospheric profiles}
Our simulations did not include the use of a general circulation model (GCM). GCM simulations at conducted at the very high irradiations and temperatures expected for the three planets would be extremely challenging in terms of model stability. \cite{Koll_Abbot2016} have shown that simulating the temperature structure of hot (up to 10 times Earth's irradiation) and dry rocky exoplanets with a GCM is only relevant if the atmosphere of high mean molecular weight is hot or thin $(p\subtxt{surf} \leq1)$. Thick atmospheres are expected to be more homogeneous due to a more efficient heat transport, smoothing out horizontal variabilities, mitigating the need for a GCM to simulate the atmospheric structure. Also, if the atmosphere is moist, such as in our steam atmosphere scenario, latent heat transport would reduce the day–night temperature gradient compared to dry atmospheres \citep{Leconte2013}. Therefore, the lack of resolving general circulation in our simulations should not have a significant impact on the temperature and mixing ratio profiles at the terminator. \\

We did not explicitly compute escape processes. Instead, we calculated thermal escape rates based on the planet’s equilibrium temperature and assumed a Bond albedo of 0.3. A simple calculation such as this one may have over or underestimated the true rate of escape. Further work would be needed to model escape from this system in detail, but we expect that an H$_2$ atmosphere would be unstable on short timescales ($<100$ Myr) under a variety of escape mechanisms, including XUV-driven photo-evaporation \citep{owen2012,tripathi2015,lopez2017,murrayclay2018}, classical Jeans escape, and thermal boil-off \citep{Owen2017}. \\ 

\subsection{Emission spectroscopy}
It is important to note that emission spectroscopy could be valuable for highly irradiated planets such as the L 98-59 system. \cite{Morley2017} showed that in the case of TRAPPIST-1b, thermal emission spectra taken in the JWST/MIRI wavelength range are more sensitive to surface pressure and planetary equilibrium temperature than transmission spectra. \cite{Lincowski2018} also showed that emission spectroscopy can be used in synergy with transmission spectroscopy to improve the discrimination between various atmospheric scenarios. However, in this work, simulations within the JWST/MIRI wavelength range are only considered for O$_2$-dominated atmospheres to constrain the detectability of O$_2$-O$_2$ CIA at 6.4 $\mu$m. These modeled atmospheres are isothermal, so they do not have a temperate structure that would allow us to see spectral features in thermal emission. 

\subsection{Noise and instrumental configurations}
\label{sub:noise}
In this study, we have placed our transmission spectrum simulations in a photon-limited scenario in which the noise would follow a white noise decay, i.e. proportional to $1 / \sqrt{X}$ or X$^{-0.5}$ with X being the number of transits. No noise floor is assumed. We therefore did not consider noise systematics that can be added to the white noise and therefore reduce (in absolute value) the X exponent. Systematics can come from the instrument, for instance due to intra-pixel gain variability \citep{Knutson2008,Anderson2011}, tracking uncertainties, etc. \cite{kreidberg2014} observations of 15 transits of GJ 1214 with HST/WFC3 have shown a noise decay following a white noise model. For 55 Cancri e, \cite{Tsiaras2016} reached 20-30~ppm precision over 25 channels in a single transit with HST/WFC3. The fact that a noise floor better than 30~ppm has not been achieved yet with HST/WFC3 does not mean that it is the precision limit of the instrument, but instead that not enough transits have been accumulated to actually reach the noise floor. This may be achieved faster with JWST, as its aperture size allows for a larger photon collecting area. \cite{Deming2009, Greene2016} have assumed 1 $\sigma$ noise floors for NIRSpec and MIRI LRS of 20 and 50~ppm, respectively, based on the current performance of HST/WFC3. However, this neglects the improvement of the detector stability and data reduction techniques. JWST's instrumental noise floors are likely smaller than that \citep{Fauchez2019,Pidhorodetska2020}, but could not be accurately estimated prior to launch. Therefore, the use of a photon limited (white noise) scenario is perhaps adequate for these HST and JWST simulations.

\section{Conclusion} 
\label{sec:conclusion}
Our investigation of the potential to detect and characterize the atmospheres of the L 98-59 planets through transmission spectroscopy indicates that 1 transit with HST/WFC3 could detect a low mean-molecular weight atmosphere on L 98-59 c and d, while 1 transit with JWST/NIRISS SOSS would allow us to begin distinguishing between a variety of possible atmospheres for each planet in the L 98-59 system. Although the planets are small and likely possess high mean molecular weight atmospheres with relatively low scale heights, we found that many molecular absorption features may be detectable with JWST/NIRSpec G395H and NIRISS SOSS in 2-26 transits. \\ 

We find that observations with HST/WFC3 could lead to a 5$\sigma$ detection of atmospheric spectral features such as CH$_4$ and H$_2$O in 1 transit for H$_2$-dominated atmospheres on L 98-59 c and d. In addition, HST/WFC can detect absorption features in a higher clear mean-molecular weight atmosphere such as one dominated by H$_2$O in 6 transits or less for L 98-59 b, c, and d. When considering the presence of an organic haze in a steam atmosphere, HST/WFC3 could rule out a featureless spectrum in 20 transits or less for each planet. However, observations with JWST/NIRISS SOSS could detect the same clear H$_2$O atmosphere in 1 transit or less for each planet, and the presence of an organic haze would only increase this to 2 transits or less for each planet. This boost in detectability is partially due to JWST's increased sensitivity, but can also be attributed to NIRSpec's extended wavelength range, where haze opacity becomes negligible in the MIR. With access to redder wavelengths, we can probe for multiple H$_2$O absorption bands, allowing us to confirm the presence of H$_2$O with greater confidence while also placing constraints on H$_2$O abundance via retrieval modeling. \\

Planets in the Venus Zone that can be spectroscopically characterized are important in the realm of comparative planetology that aims to characterize the conditions of a post-runaway atmosphere. JWST has a strong ability to detect the presence of an atmosphere dominated by CO$_2$, as it possesses numerous absorption bands from the near- through the mid-IR, such as the 2.0, 2.7, 4.3, and 15 $\mu$m bands. This allows for a 5$\sigma$ detection of a clear Venus-like atmosphere with 16, 26, and 17 transits for L 98-59 b, c, and d, respectively. The addition of Venus-like H$_2$SO$_4$ clouds would increase the number of transits required for the same detection to 37, 80, and 57 transits for L 98-59 b, c, and d, respectively. However, a Venus-like planet that has lost its residual hydrogen would not be able to form H$_2$SO$_4$ clouds, creating an opportunity for observing stronger features. Although CO$_2$ makes for a strong indicator of an atmosphere in our simulations, it is a weak discriminant of any specific atmospheric state. Other molecules such as O$_2$, O$_3$, and H$_2$O may be detectable with JWST, aiding in the distinction between the suite of atmospheres explored within this work. \\ 

Highly irradiated planets such as those of the L 98-59 system could have a desiccated atmospheric composition, such as one that is dominated by O$_2$, as a result of major ocean loss during an extended runaway greenhouse phase. A desiccated planet that is rich in abiotic O$_2$ would be expected to form O$_3$ from the photochemical processing of O$_2$, meaning that the direct detection of O$_3$ absorption could be another key indicator of this planetary state. This reaction is temperature dependent, where hot planetary atmospheres (T$>$300K) significantly suppress O$_3$ formation. In the JWST/NIRISS SOSS (0.6-2.8 $\mu$m) and MIRI LRS wavelength ranges (5.0-12.0 $\mu$m), the O$_2$-O$_2$ CIA band at 1.27 $\mu$m provides the strongest S/N compared to all other O$_2$ bands, allowing for a 5$\sigma$ detection with 20 transits or less (1 bar) and 7 transits or less (10 bar) for all three planets. However, when considering planetary equilibrium temperatures of 200K, the reduced scale height of the absorption features increases the number of transits required for detection to 55, 57, and 31 transits (1 bar) and 28, 57, and 31 transits (10 bar) for L 98-59 b, c, and d; respectively. At these lower temperatures, we see an increase in O$_3$ formation, but the O$_3$ absorption features between 2-5 $\mu$m and at 9.7 $\mu$m are too weak to allow for a confident detection. As MIRI LRS calls for significantly more transits to make a confident detection of O$_2$, we find that NIRSpec and NIRISS are JWST's optimal instruments to conduct transmission spectroscopy measurements of the L 98-59 planets. We also note that the detection of H$_2$O in the atmospheres of the L 98-59 planets may help to constrain evolutionary scenarios.  High O$_2$ atmospheres for planets that exited the pre-main-sequence with their atmospheres and interiors completely desiccated, however, will have no water to detect, making water in an O$_2$ dominated atmosphere a potentially detectable discriminant of incomplete desiccation or outgassing from the planetary interior \citep{Lustig-Yaeger2019}. 

\acknowledgments
D.P. and E.W.S acknowledge support from the NASA Astrobiology Institute's Alternative Earths team funded under Cooperative Agreement Number NNA15BB03A and the Virtual Planetary Laboratory, which is a member of the NASA Nexus for Exoplanet System Science, and funded via NASA Astrobiology Program Grant No. 80NSSC18K0829. S.E.M acknowledges support from NASA Earth and Space Science Fellowship grant 80NSSC18K1109. E.A.G. thanks the LSSTC Data Science Fellowship Program, which is funded by LSSTC, NSF Cybertraining Grant \#1829740, the Brinson Foundation, and the Moore Foundation; her participation in the program has benefited this work. Goddard affiliates acknowledge support from the GSFC Sellers Exoplanet Environments Collaboration (SEEC), which is supported by NASA’s Planetary Science Division’s Research Program.

\bibliography{References.bib}

\end{document}